\documentclass[twoside]{article}
\usepackage{atmp}
\usepackage{amsmath}%
\usepackage{amsfonts}%
\usepackage{times}
\usepackage{amsthm} 
\usepackage{epsfig}                
\newcommand{\bchange}
{\marginpar{$\Downarrow\Downarrow$}}   
\newcommand{\echange}
{\marginpar{$\Uparrow\Uparrow$}}       
\newcommand{\comma}{\: ,}              
\newcommand{\period}{\: .}             
\newcommand{\Proof}{\noindent{\em Proof: }}               
\newcommand{\QED}{\qed}
\newcommand{\eps}{{\varepsilon}}        
\newcommand{\vphi}{{\varphi}}           
\newcommand{\om}{\omega} \newcommand{\Om}{\Omega}
\newcommand{\la}{\langle} \newcommand{\ra}{\rangle}
\newcommand{\ol}{\overline} 
\newcommand{\one}{\mathbf{1}}

\newcommand{\cB}{\mathcal{B}}
\newcommand{\cD}{\mathcal{D}}

\newcommand{\cF}{\mathcal{F}}

\newcommand{\cH}{\mathcal{H}}

\newcommand{\cM}{\mathcal{M}}         
\newcommand{\cO}{\mathcal{O}}         

\newcommand{\cT}{\mathcal{T}}

\newcommand{\RR}{\mathbb{R}}            
\newcommand{\NN}{\mathbb{N}}            
\newcommand{\CC}{\mathbb{C}}            
\newcommand{\PP}{\mathbb{P}}            
           
\newcommand{\ZZ}{\mathbb{Z}} 
\renewcommand{\SS}{\mathbb{S}}
\newcommand{\uf}{{\underline f}}

\newcommand{\uvphi}{{\underline{\varphi}}}

\newcommand{\hG}{\widehat{G}}            

\newcommand{\hrho}{\hat{\rho}}
\newcommand{\hpsi}{\hat{\psi}}
\newcommand{\hvphi}{\hat{\varphi}}

\newcommand{\tH}{\widetilde{H}}         
\newcommand{\tQ}{\widetilde{Q}}         
\newcommand{\tW}{\widetilde{W}}         
\newcommand{\te}{\tilde{e}}             
\newcommand{\tf}{\tilde{f}}
\newcommand{\tk}{\tilde{k}}

\newcommand{\tchi}{\tilde{\chi}}
\newcommand{\tpsi}{\tilde{\psi}}

\newcommand{\vk}{{\vec{k}}}

\newcommand{\vx}{{\vec{x}}}

\newcommand{\be}{\begin{equation}}
\newcommand{\ee}{\end{equation}}

\newcommand {\field}[1]{\mathbb{#1}}

\newcommand {\R}{\field R}

\newcommand{\rIm}{\mathrm{Im}}
\newcommand{\rRe}{\mathrm{Re}}               
\newcommand{\Ran}{\mathrm{Ran}}              
\newcommand{\spec}{\sigma}                   

\newcommand{\cirS}{\mathop{\bigcirc\kern -.73em {\scriptstyle\mathrm{S}}}}

\newcommand{\supp}{\mathrm{supp}} 
\newcommand{\dom}{\mathrm{dom}}

\newcommand{\chf}{\mathbf{1}} 
\newcommand{\el}{{el}} 
\newcommand{\rad}{{f}} 
 
\newcommand{\hel}{H_\el} 
\newcommand{\thel}{\widetilde{H}_\el} 
\newcommand{\hf}{H_\rad} 
\newcommand{\hfL}{H_\rad^{(\Lambda)}} 

\newcommand{\DV}{\Delta V}
\newcommand{\gs}{\Phi_{\mathrm{gs}}}
\newcommand{\Qinf}{Q^{\mathrm{inf}}}
\newcommand{\Qsup}{Q^{\mathrm{sup}}}

\newcommand{\Rem}{\mathrm{Rem}}
\renewcommand{\thesection}
{\Roman{section}}                      
\renewcommand{\theequation}
{\thesection.\arabic{equation}}        
\newcommand{\secct}[1]{\section{#1}
\setcounter{equation}{0}}              
\newtheorem{hypothesis}{Hypothesis}
\newtheorem{theorem}{Theorem}[section]         
\newtheorem{lemma}[theorem]{Lemma}             
\theoremstyle{plain}
\begin{document}
\bibliographystyle{plain}
\copyrightnotice{2001}{5}{969}{999}
\setcounter{page}{969}
\title{Mathematical Analysis of \\ the Photoelectric Effect}
\author{ 
Volker Bach, \ 
Fr{\'e}d{\'e}ric Klopp, \ and  \  
Heribert Zenk }
\address{FB Mathematik; Johannes Gutenberg-Universit{\"a}t; 
D-55099 Mainz; Germany; }
\addressemail{vbach@mathematik.uni-mainz.de} 
\address{LAGA, Institut Galil{\'e}e; Universit{\'e} Paris-Nord; 
F-93430 Villetaneuse; France }
\addressemail{klopp@math.univ-paris13.fr}
\address{
FB Mathematik; Johannes Gutenberg-Universit{\"a}t; 
D-55099 Mainz; Germany;} 
\addressemail{zenk@mathematik.uni-mainz.de 
}
\date{}
\begin{abstract}
  We study the photoelectric effect on the example of a simplified
  model of an atom with a single bound state, coupled to the quantized
  electromagnetic field.
  
  For this model, we show that Einstein's prediction for the
  photoelectric effect is qualitatively and quantitatively correct to
  leading order in the coupling parameter. More specifically,
  considering the ionization of the atom by an incident photon cloud
  consisting of $N$ photons, we prove that the total ionized charge is
  additive in the $N$ involved photons. Furthermore, if the photon
  cloud is approaching the atom from a large distance, the kinetic
  energy of the ejected electron is shown
  to be given by the difference of the photon energy of each single
  photon in the photon cloud and the ionization energy.
\vspace{2mm}


\vspace{2mm}

\end{abstract}
\pagestyle{myheadings}
\markboth{Bach, Klopp, and Zenk}{Mathematical Analysis of the Photoelectric Effect}

\thispagestyle{empty}

\newpage

\secct{Introduction} \label{sec-I}
%
The photoelectric effect was discovered in increasingly precise
experiments by Hertz \cite[1887]{Hertz1887}, Hallwachs
\cite[1888]{Hallwachs1888}, Lenard 
\cite[1902]{Lenard1902}\footnote{In some physics textbooks, e.g.,
  \cite{HallidayResnick1978}, Lenard is not mentioned, and
  it seems that in 1905, Einstein derived his famous
  theory from nothing but a Gedankenexperiment.}, and Millikan
\cite[1916]{Millikan1916a,Millikan1916b}.

It was observed that, when light is incident on a metal surface,
electrons are ejected from the surface. The striking fact about this
phenomenon is the seemingly odd dependence of the maximal kinetic
energy $T_{max}$ of the ejected electrons on the frequency of the
light and its independence of the light intensity.
The latter contradicts the principles of classical physics, and
in 1905, Einstein suggested an explanation of this phenomenon
\cite[1905]{Einstein1905} which explicitly involves the
{\em quantum} nature of electromagnetic radiation.
He found that
\begin{equation} \label{eq-I-1}
T_{max} \ = \ h \, \nu \; - \; \Delta E \comma
\end{equation}
provided that the frequency $\nu$ of the light times Planck's
constant $h$, i.e., the photon energy $h \nu$
is larger than the (material dependent) work function $\Delta E$.
Conversely, if
\begin{equation} \label{eq-I-2}
h \, \nu \ < \ \Delta E \comma
\end{equation}
then no electrons leave the metal surface.  Our ultimate goal is the
derivation of Einstein's predictions, Eqs.~(\ref{eq-I-1})--(\ref{eq-I-2}), 
from first principles of quantum mechanics and quantum field theory.

In the present paper, we analyze a simplified model which is far from
the appropriate model for a metal interacting with electromagnetic
radiation and can, at best, be regarded as a caricature of a hydrogen
atom interacting with the radiation field. Yet, it contains many of
the mathematical difficulties we expect to encounter in the analysis of
a more realistic model, and we prove Eqs.~(\ref{eq-I-1})--(\ref{eq-I-2}) 
for this simplified model. Our emphasis lies in the following aspects:
\begin{itemize}
\item
Given the model as described in Sect.~\ref{subsec-I-1}, below,
our derivation is mathematically rigorous, and no unjustified
approximations are used. To our knowledge, the present paper 
is the first to treat the photoelectric effect with mathematical 
rigor. 

We draw from many facts about nonrelativistic
quantum electrodynamics which have been previously established in
\cite{BachFroehlichSigal1998a,BachFroehlichSigal1998b,BachFroehlichSigal1999,BachFroehlichSigal2000}.

 As we propose to study the \emph{charge transported to infinity} (see
Sect.~\ref{subsec-I-2}) which involves the asymptotics of the unitary
time evolution operator $e^{-itH_g}$, as $t \to \infty$, our results
can be viewed as part of scattering theory for models of 
nonrelativistic quantum electrodynamics. Results in this context, but
on other aspects can be found in
\cite{AmreinBoutetdeMonvelGeorgescu1996,DerezinskiGerard1997,FroehlichGriesemerSchlein2001,Gerard1996,GriesemerLiebLoss2001,Schlein1999}.
\item 
While our model for the particle system, i.e., the metal or
atom, is a crude model involving only a single, spinless electron,
the particle system is coupled to a {\em quantized} scalar field.
(The difference between the quantized [vector] electromagnetic and a
quantized scalar field is irrelevant, for the scope of this work.
For certain other facts in nonrelativistic quantum electrodynamics,
however, this difference is crucial, see, e.g., 
\cite{BachFroehlichSigal1999}.)
\item
Our model describes a single atom rather than a metal
or even a gas interacting with the electromagnetic field. Thus our
derivation of Eqs.~(\ref{eq-I-1})--(\ref{eq-I-2}) shows that
the photoelectric effect is not a collective, statistical phenomenon, 
visible only if \emph{many} particles or \emph{many} photons are involved.
Most texts on laser theory or quantum optics, e.g., 
\cite{Eastham1986,Haken1981}, immediately proceed to 
a statistical description of both the metal or gas of atoms, say,
and the photon field, and the question, whether this is really
necessary or merely a matter of mathematical convenience, is left open.
\item 
Within our framework, we prove Eqs.~(\ref{eq-I-1})--(\ref{eq-I-2})
to be correct in leading nonvanishing order in the coupling
parameter $g$ which, in appropriate units, equals 
$\sqrt{2\pi} \, \alpha^{3/2}$, where $\alpha \approx 1/137$ is the 
fine structure constant. More precisely, given the interacting atom at
rest plus an incident photon cloud consisting of $N$ photons, we show
that to leading order the contribution to the charge ejected from the
atom is \emph{additive} in each photon and \emph{independent}
of all other photons of the incident photon cloud. Moreover, we prove
that this contribution is in accordance with
(\ref{eq-I-1})--(\ref{eq-I-2}) in case that the photons are in an
incoming scattering state. In fact, the leading order contribution 
to the ejected charge resembles the first term in the Born series for the
T-Matrix (see, e.g., \cite{MR80m:81085}).  
\item 
While it is customary to restrict the analysis of the
photoelectric effect to a single photon scattering off the atom or
metal, we point out that it is important to consider more than one
photon, $N \geq 2$, because for a photon state consisting of a
single photon only, total energy and energy of each single photon
involved agree. Hence, if studying a single photon state, it is
impossible to say whether the ejection of electrons is proportional
to the total energy of the photon cloud or depends on the maximal
energy of all photons in the photon cloud.
\end{itemize}
Our paper is organized as follows.
In Sect.~\ref{subsec-I-1} below we introduce the mathematical model
for the photoelectric effect. This includes a precise description
of the atom, the photons, and their interaction in terms of a 
semibounded, selfadjoint generator of the dynamics, the Hamiltonian.

In Sect.~\ref{subsec-I-2} we then describe our main results. 
In Subsects.~\ref{subsubsec-I-2-1}--\ref{subsubsec-I-2-2},
we introduce the main quantities dealt with in this paper, the charge 
transported to infinity and photon clouds, and in 
Subsect.~\ref{subsubsec-I-2-3}, we derive the asymptotics of the former
to leading order in the coupling constant. A limit of monochromatic
light is discussed in Subsect.~\ref{subsubsec-I-2-4}, and in
Subsect.~\ref{subsubsec-I-2-5}, we compare our methods and results
to those derived or used in other papers on scattering theory. 

The proof of the theorems in 
Sect.~\ref{subsec-I-2} are given in detail in Sects.~\ref{sec-II}
and \ref{sec-IV}.

Finally, our paper contains two appendices.
In Appendix~\ref{sec-A} we construct a Bogoliubov transformation
that eliminates a single, arbitrary matrix element in the interaction,
and in Appendix~\ref{sec-B} we show that bound or negative energy states
do not contribute to the transported charge. 

\vspace{3mm}

\noindent \textbf{Acknowledgement:}
We thank S.~De Bi{\`e}vre, J.~Fr{\"o}hlich, Ch.~Gerard, M.~Griesemer, 
V.~Ko\-stry\-kin, T.~Paul, B.~Schlein, R.~Schrader, I.~M.~Sigal, 
H.~Spohn, and S.~Teufel for helpful discussions and remarks.

We gratefully acknowledge the support of our project 
by the TMR network ERBFMRXCT960001 of the European Union
and the Gerhard Hess-grant Ba~1477/3-1 of the german science
foundation DFG.


\subsection{Mathematical Model for the Photoelectric Effect} 
\label{subsec-I-1}
%
Our goal is to analyze the photoelectric effect on the example of
an atom coupled to the quantized radiation field.
The mathematical model for our analysis of the photoelectric effect
is a variant of the {\em standard model of nonrelativistic quantum
electrodynamics} as given in \cite{BachFroehlichSigal1998a} which
we briefly recall and adapt to our problem at hand. 

\subsubsection{Atom with a Single Bound State}
\label{subsubsec-I-1-1}
For the atom to be described we make the simplifying assumptions that
it consists of a single, spinless electron bound in a potential well
which admits exactly one bound state of energy $e_0 <0$. More
specifically, we assume the Hamiltonian generating the dynamics of the
electron to be given in diagonal form as
\begin{equation} \label{eq-I-3}
\hel \ = \ -\Delta \oplus e_0 \ = \ 
\left( \begin{array}{cc} -\Delta & 0\\ 0 & e_0 \end{array} \right) \comma 
\end{equation}
acting on 
\begin{equation} \label{eq-I-4}
\cH_\el \ := \ L^2(\RR^3) \oplus \CC 
\ = \ \cH_{ac}(\hel) \oplus \cH_{d}(\hel) \comma
\end{equation}
and $\cH_{sc}(\hel) = \{0\}$. We further denote by
\begin{equation} \label{eq-I-5}
P_c \ := \ 
\left( \begin{array}{cc} \one & 0\\ 0 & 0 \end{array} \right) 
\hspace{5mm} \mbox{and} \hspace{5mm}
P_d \ := \ P_c^\perp \ = \ 
\left( \begin{array}{cc} 0 & 0\\ 0 & 1 \end{array} \right) 
\end{equation}
the projections onto the continuous subspace and the discrete subspace
(of dimension one), respectively. The Hamiltonian $\hel$ can be
derived from a Schr{\"o}dinger operator $-\Delta - V(x)$ 
with a short-range potential $V$. An appropriate choice of $V$ 
guarantees that $-\Delta - V(x)$ has a single bound state only,
and then $-\Delta - V(x)$ and $H_\el$ are unitarily equivalent,
as can be seen by conjugating $-\Delta - V(x)$ with the wave
operators.

\subsubsection{Photons}
\label{subsubsec-I-1-2}
We couple the atom described above to the quantized photon field
which, for notational convenience, is assumed to be scalar. For our
study of the photoelectric effect, the difference is not relevant.
The Hilbert space $\cF$, carrying the photon degrees of freedom is the
bosonic Fock space $\cF = \cF_b [ L^2( \RR^3 )]$ over the one-photon
Hilbert space $L^2( \RR^3 )$, i.e.,
\begin{equation} \label{eq-I-6}
\cF \ = \  \bigoplus_{n=0}^\infty \, \cF^{(n)} \comma
\end{equation}
where $\cF^{(n)}$ is the state space of all $n$-photon states,
the $n$-photon sector. The vacuum sector, $\cF^{(0)}$, is 
one-dimensional and spanned by the normalized vacuum vector, 
$\Om$, i.e., $\cF^{(0)} := \CC \, \Om$. 
For $n \geq 1$, the $n$-photon sector is the subspace of
$L^2[(\RR^3)^n]$ containing all totally symmetric vectors.
The Hamiltonian on $\cF$ representing the energy of the free photon 
field is given by
\begin{equation} \label{eq-I-8}
\hf \ = \ \int d^3k \, \om(k) \, a^*(k) a(k) \comma
\end{equation}
where $\om(k) := |k|$ is the photon dispersion, and 
$a^*$, $a$ are the usual standard creation- and annihilation
operators on $\cF$ representing the canonical commutation relations,
$[ a(k), a(k') ] $ \newline\noindent $= [ a^*(k), a^*(k') ] = 0$,
$[ a(k), a^*(k') ] = \delta(k-k')$,  $a(k) \Omega =0$,
in the sense of operator-\-valued distributions.

\subsubsection{The Atom-Photon System}
\label{subsubsec-I-1-3}
The Hilbert space of states of the atom-photon system
is the tensor product space
\begin{equation} \label{eq-I-10}
\cH \ := \ \cH_\el \otimes \cF \comma
\end{equation}
and its dynamics is generated by the Hamiltonian
\begin{equation} \label{eq-I-11}
H_g \ := \ H_0 \; + \; g \, W \comma
\end{equation}
where
\begin{equation} \label{eq-I-12}
H_0 \ := \ \hel \otimes \one_f \; + \; \one_\el \otimes \hf 
\end{equation}
is the non-interacting Hamiltonian, $0< g \ll 1$ is a small coupling
parameter, and 
\begin{equation} \label{eq-I-13}
W \ := \ \int d^3k 
\Big\{ G(k) \otimes a^*(k) \; + \; G^*(k) \otimes a(k) \Big\}
\end{equation}
is the interaction operator. Here, $G \in L^2[ \RR^3; \cB(\cH_\el)]$ is a
square-integrable function with values in the bounded operators on
$\cH_\el$, given by
\begin{equation} \label{eq-I-14}
G(k) \ := \ 
\left( \begin{array}{cc} 
B(k)            & p_\uparrow(k) \\ 
p_\downarrow(k) & 0             \\   \end{array} \right) \comma 
\end{equation}
for $k \in \RR^3$, a.e. For the formulation of assumptions about $G$
it is convenient to introduce 
\begin{equation} \label{eq-I-14.1}
J_{K,\gamma} (k) \ := \ 
\max_{|\alpha| \leq K} \big\| \partial_k^\alpha G(k) \big\|
\, + \, \big\| (|x|^\gamma \oplus 1) \, G(k) \big\|
\, + \, \big\| G(k) \, (|x|^\gamma \oplus 1) \big\| \period
\end{equation}
Note that, for $J_{K,\gamma} (k) < \infty$, we necessarily assume
a decay of the coupling matrix $G(k)$ as least as $|x|^{-\gamma}$,
as $x \to \infty$. This assumption is not satisfied, e.g., if
we consider electrons minimally coupled to the quantized radiation
field. 
Namely, in the case of minimal coupling, the linear part of the
interaction operator is of the form (\ref{eq-I-13}), with
\begin{equation} \label{eq-I-14.1.1}
B_{m.c.}(k) \ = \ 
\frac{ \kappa(k) \, e^{i \vk \cdot \vx} }{i \: |k|^{1/2}} \, 
\vec{\eps}(k) \cdot \vec{\nabla}_x \comma
\end{equation}
where $k = (\vk, \sigma) \in \RR^3 \times \ZZ_2$, $\kappa$ is a smooth
function of rapid decay, as $|\vk| \to \infty$, serving as an
ultraviolet cutoff, and $\vec{\eps}(\vk, \pm 1) \perp \vk$ are two
transversal, normalized polarization vectors. Besides the lack of
decay of $B_{m.c.}(k)$, as $|x| \to \infty$, $B_{m.c.}(k)$ is not
bounded. This is, however, only a minor complication. 
The extension of our
results to this case of main physical interest will be the subject of
a forthcoming paper.

Further note that
\begin{eqnarray} \label{eq-I-14.2}
\lefteqn{
J_{K,\gamma} (k) \ \leq \ J_{K,\gamma}' (k) \ := \ 
\max_{|\alpha| \leq K} \big\| \partial_k^\alpha B(k) \big\|
\, + \, \big\| \: |x|^\gamma  \, B(k) \big\|
\, + \, \big\| B(k) \, |x|^\gamma \: \big\| 
}
\\ \nonumber & & 
\, + \,
\max_{|\alpha| \leq K} \big\| \partial_k^\alpha p_\uparrow(k) \big\|
\, + \,
\max_{|\alpha| \leq K} \big\| \partial_k^\alpha p_\downarrow(k) \big\|
\, + \, \big\|  \: |x|^\gamma  \, p_\uparrow(k) \big\|
\, + \, \big\| p_\downarrow(k) \, |x|^\gamma \: \big\| \comma
\hspace{2mm}
\end{eqnarray}
and, conversely, $J_{K,\gamma}' (k) \leq 3 J_{K,\gamma}(k)$.
We shall make use of the following Hypothesis throughout the paper.
\begin{hypothesis} \label{H-1}
There is an integer $K > 1$ and a real number $\gamma > 3/2$
such that $G \in C^K[ \RR^3; \cB(\cH_\el)]$ is $K$
times differentiable and satisfies the following estimate,
\begin{equation} \label{eq-I-14.3} 
\int \Big( 1 + \omega(k)^{-1} \Big) \, | J_{K,\gamma} (k) |^2
\, d^3k \ \leq \ 1 \period 
\end{equation}
\end{hypothesis}
We remark that, although the requirement (\ref{eq-I-14.3}) with
fractional derivatives of $G$, for real $K>1$, is presumably
sufficient, we do not try to optimize our result in this respect and
work with classical derivatives in this paper.

Since, for a.e. $k \in \RR^3$, 
\begin{equation} \label{eq-I-15} 
p_\uparrow (k) : \CC \; \to \; L^2(\RR^3) 
\hspace{5mm} \mbox{and} \hspace{5mm}
p_\downarrow (k) : L^2(\RR^3) \; \to \; \CC \comma
\end{equation}
there exist $\rho(\, \cdot \, , k), \eta(\, \cdot \, k) \in L^2(\RR^3)$, 
such that
\begin{equation} \label{eq-I-16} 
\big[ p_\uparrow (k) z \big](x) \: = \: z \, \eta(x, k)  
\hspace{3mm} \mbox{and} \hspace{3mm}
p_\downarrow (k) \psi \: = \:  \la \rho(\, \cdot \, , k) | \psi \ra 
\: = \: \int \ol{\rho(x, k)} \psi(x) \, d^3x \comma  
\end{equation}
for all $z \in \CC$, $\psi \in L^2(\RR^3)$, and 
$(x, k) \in \RR^3 \times \RR^3$, a.e. 
Furthermore, in many applications 
$B(k) \in \cB[L^2(\RR^3)]$ acts in the Schr{\"o}dinger
representation as a multiplication operator. That is, there
is a function $M \in L^2(\RR^3 \times \RR^3)$ such that
\begin{equation} \label{eq-I-17} 
\big[ B(k)\psi \big](x) \ = \  M(x, k) \, \psi(x) \comma
\end{equation}
for $(x, k) \in \RR^3 \times \RR^3$, a.e. For instance, in case
of the dipole approximation, 
\begin{equation} \label{eq-I-17.1}
M_{dip}(x, k) \ = \ 
\kappa(\vx/R) \, \kappa(k) \, |k|^{1/2} \: \vec{\eps}(k) \cdot \vx \comma
\end{equation}
using the same notation as for $B_{m.c.}$, above, and additionally
a spatial cutoff $\kappa(\vx/R)$ at length scale $R \gg 1$ which 
should be chosen large, compared to atomic length scales. 
We remark that, for a given function $M$, a physically natural choice
for $\eta$ and $\rho$ is 
\begin{equation} \label{eq-I-22}
\eta(x,k) \ := \ \rho(x,k) \ := \ M(x,k) \, \vphi_\el(x) \comma
\end{equation}
where $\vphi_\el \in L^2(\RR^3)$, $\| \vphi_\el \| =1$, 
is the normalized wave function of the atomic bound state. 
Note that the special form (\ref{eq-I-14}) of $G(k)$ implies that
\begin{equation} \label{eq-I-23.1}
(P_d \otimes \one) \; W \; (P_d \otimes \one) \ = \ 0 \comma
\end{equation}
since the lower right matrix element of $G(k)$ vanishes. 

We remark that if the electron Hamiltonian is a Schr{\"o}dinger operator,
as described below Eq.~(\ref{eq-I-5}), a \emph{single} diagonal matrix 
element can always be ``gauged away'' by means of a Bogoliubov 
transformation. The construction of this Bogoliubov transformation is
given in Appendix~\ref{sec-A}. That is,
for our model with a \emph{single} atomic bound state, we do not
loose any generality by assuming (\ref{eq-I-23.1}). 
As (\ref{eq-I-23.1}) is an important assumption for the present
paper, we point out that, in case the atom has more than one bound state, 
$P_d$ would be the orthogonal projection onto these, and (\ref{eq-I-23.1})
would imply, that the field does not couple different bound states.

\subsubsection{Selfadjointness, Semiboundedness, and Binding}
\label{subsubsec-I-1-4}
Next, we discuss selfadjointness, semiboundedness and
existence of a ground state (binding) of the Hamiltonian $H_g$ of the
system. We basically invoke the theorems and methods for their proof
established in \cite[Sects.~II and III]{BachFroehlichSigal1998a}.

Assuming Hypothesis~\ref{H-1}, selfadjointness and semiboundedness,
for any value of the coupling parameter $g>0$, are a simple consequence
of standard Kato perturbation theory; we appeal to
\cite[Cor.~I.7 and Lemma~I.5]{BachFroehlichSigal1998a} which yield that $H_g$
is selfadjoint on its natural domain, $\dom[H_g] = \dom[H_0]$,
and that the ground state energy is given by
\begin{equation} \label{eq-I-29} 
E_0 \ := \ \inf\spec[H_g] \ \geq \ e_0 + \cO(g^2) \period
\end{equation}
In the present paper we additionally assume that $E_0$ is
an eigenvalue of $H_g$ and that the corresponding eigenvectors have
a large component in the vacuum sector $\cF^{(0)}$. More precisely, 
we require
\begin{hypothesis} \label{H-2} 
The Hamiltonian $H_g$ possesses a (normalizable) ground state 
$\gs \in \cH$, $\| \gs \|=1$, i.e., $E_0$ is an eigenvalue with
corresponding eigenvector $\gs$,
\begin{equation} \label{eq-I-33} 
H_g \, \gs \ = \ E_0 \, \gs \period
\end{equation}
Moreover, denoting 
$P_\Om := |\Om\ra\la\Om| \equiv \one \otimes |\Om\ra\la\Om|$,
the ground state $\gs$ obeys 
\begin{equation} \label{eq-I-34} 
\big\| P_\Om^\perp \; \gs \big\| \ \leq \ C \, g \comma
\end{equation}
for some constant $C < \infty$.
\end{hypothesis}
The existence (\ref{eq-I-33}) of a ground state $\gs$ and the overlap
bound (\ref{eq-I-34}) is proved in 
\cite[Thm.~I.1]{BachFroehlichSigal1998a} under the assumption of a somewhat
stronger bound than Eq.~(\ref{eq-I-14.3}) in Hypothesis~\ref{H-1}, namely,
\begin{equation} \label{eq-I-29.1} 
\int \Big( 1 + \omega(k)^{-2} \Big) \, | J_{0,\gamma} (k) |^2
\, d^3k \ \leq \ 1 \period 
\end{equation}
In this paper, we do not use this estimate but only its consequence
in form of Hypothesis~\ref{H-2}.

We remark that the noninteracting Hamiltonian $H_0$ has the unique
(non-de\-ge\-ne\-rate) ground state $0 \oplus \Omega \in \cH$ corresponding
to its ground state energy $e_0$, and we note in parentheses that, 
by the results of \cite{BachFroehlichSigal1998a}, Eq.~(\ref{eq-I-34}) 
holds for {\em any} ground state of $H_g$. Consequently, the 
interacting ground state $\gs$ of $H_g$ is unique, 
provided $g>0$ is sufficiently small.

In Lemma~\ref{lem-IV-1} we strengthen (\ref{eq-I-34}) and
show that, for any $\alpha \geq 1$ and any $\Lambda < \infty$,
\begin{equation} \label{eq-I-34a} 
\big\| (\hfL+\one)^\alpha \, P_c \, \gs \big\| \: + \:
\big\| (\hfL+\one)^\alpha \, P_\Om^\perp \, \gs \big\| 
\ \leq \ \cO(g) \period
\end{equation}
Here, $\hfL$ denotes a free photon Hamiltonian acting only
on photon states with energy less than $\Lambda < \infty$,
\begin{equation} \label{eq-I-34b} 
\hfL \ = \ \int d^3k \, \om_\Lambda(k) \, a^*(k) a(k) 
\ = \ \int_{\{\om(k) \leq \Lambda\}} d^3k \, \om(k) \, a^*(k) a(k) 
\comma
\end{equation}
where $\om_\Lambda(k) := \om(k) \, \chf_{ \{\om(k) < \Lambda\} }$.

\subsection{Main Results and Discussion} 
\label{subsec-I-2}
%
\subsubsection{Charge transported to Infinity}
\label{subsubsec-I-2-1}
Having introduced the quantum mechanical framework in terms
of Hilbert spaces and Hamiltonians and established some basic
facts such as selfadjointness, semiboundedness, and binding,
we proceed to defining the {\em charge transported to infinity}
or simply the {\em transported charge}. To this end, we
introduce the projection $F_R$ onto the functions with
support outside the ball of radius $R>0$. More precisely,
\begin{equation} \label{eq-I-35} 
F_R \ := \ 
\left( \begin{array}{cc} 
\chf_{ \{ |x| \geq R \} } & 0 \\ 0 & 0 \\ \end{array} \right) 
\, \otimes \, \one_f \comma
\end{equation}
where $\chf_{ \{ |x| \geq R \} } := \chf_{\RR^3 \setminus B(0,R)}[x]$.
Similarly, we introduce the projection $T_\cT$ onto the particle
states with momentum in a measurable set $\cT \subset \RR^3$, i.e., 
the functions whose Fourier transform is supported in $\cT$. That is,
\begin{equation} \label{eq-I-35.1} 
T_\cT \ := \ 
\left( \begin{array}{cc} 
\chf_{ \{  p \in  \cT \} } & 0 \\ 0 & 0 \\ \end{array} \right) 
\, \otimes \, \one_f \period
\end{equation}
where $\chf_{ \{  p \in  \cT \} }$ acts as a Fourier multiplier with
$\chf_\cT [p]$.  Given a state $\Psi \in \cH$, the corresponding
(least and most) transported charges with momentum in 
$\cT \subset \RR^3$ are defined to be
\begin{eqnarray} \label{eq-I-36a} 
\Qinf_\cT (\Psi) & := & \liminf_{R \to \infty} \: \liminf_{t \to \infty}
\big\| \, T_\cT \, F_R \: e^{-it H_g} \: \Psi \, \big\|^2 \comma
\\ \label{eq-I-36b}
\Qsup_\cT (\Psi) & := & \limsup_{R \to \infty} \: \limsup_{t \to \infty}
\big\| \, T_\cT \, F_R \: e^{-it H_g} \: \Psi \, \big\|^2 \comma
\end{eqnarray}
In case that $\cT = \RR^3$, we write $\Qinf(\Psi) := \Qinf_\cT (\Psi)$
and $\Qsup(\Psi) := \Qsup_\cT (\Psi)$.  Interpreting $\Psi$ as an
initial state for $t=0$, the transported charge $\Qinf(\Psi)$
measures the amount of its mass that is definitely transported away
from the atom as time evolves, eventually. If $\cT \subseteq \RR^3$
is a small ball, then $Q_\cT (\Psi)$ additionally filters out the part
of the state with momentum in $\cT$.

While it would be desirable, of course, to define only
one type of transported charge, namely, 
$Q_\cT (\Psi) := \lim_{R \to \infty} \lim_{t \to \infty}
\| T_\cT \, F_R \, e^{-it H_g} \, \Psi \|^2$, 
our methods described below do not allow us to prove the existence of 
such a limit -- not even for the restricted class of initial states 
of the form (\ref{eq-I-37}). Nevertheless, 
$\Qinf_\cT(\Psi)$ and $\Qsup_\cT (\Psi)$ agree to leading order in
the coupling constant $g$, as we demonstrate below. In fact, 
our main objective is the determination of the transported charges
$\Qinf_\cT(\Psi)$ and $\Qsup_\cT (\Psi)$ to leading order 
in $g$. Finally, we remark that it is important to observe the order of the
two limits $R \to \infty$ and $t \to \infty$, as interchanging these
limits would, indeed, yield the trivial result
$\lim_{t \to \infty} \lim_{R \to \infty} \| F_R \, e^{-it H_g} \, \Psi
\|^2 =0$, for all $\Psi \in \cH$. 

As we prove in Appendix \ref{sec-B},
\begin{eqnarray} \label{eq-I-36c} 
\lim_{R\to \infty} \sup_{t>0} \big\| F_R \, \one_{pp}(H_g) \, 
   e^{-itH_g} \, \Psi \big\| 
& = & 0 \comma
\\ \label{eq-I-36d}
\lim_{R\to \infty}  \sup_{t>0} \big\| F_R \, \one_{\RR_0^-}(H_g) \,  
   e^{-itH_g} \, \Psi \big\| 
& = & 0 \comma
\end{eqnarray}
for all $\Psi \in \cH$, so the transported charge of all bound states
$\Psi \in \Ran \one_{pp}(H_g)$ and of all states
$\Psi \in \Ran \one_{\RR_0^-}(H_g)$ of negative total energy
vanishes. 

\subsubsection{Ground State and Photon Cloud}
\label{subsubsec-I-2-2}
Our choice for the initial state $\Psi$ is of the form
\begin{equation} \label{eq-I-37} 
\Psi \ := \ A(\tau, \uf) \, \gs \comma
\end{equation}
where 
\begin{equation} \label{eq-I-38a} 
A(0, \uf) \ \equiv \ A(\uf) \ := \ 
\one_\el \, \otimes \, a^*(f_1) \, a^*(f_2) \, \cdots \, a^*(f_N) \comma
\end{equation}
and
\begin{equation} \label{eq-I-38b} 
A(\tau, \uf) \ := \ e^{-i\tau H_g} \, e^{i\tau H_0} \,  A(\uf) 
\, e^{-i\tau H_0} \, e^{i\tau H_g} \period 
\end{equation}
The operator $A(\tau, \uf)$ is called a {\em photon cloud} and
represents $N$ photons with corresponding smooth orbitals $\uf = (f_1,
f_2, \ldots, f_N)$, $f_j \in C_0^\infty(\RR^3 \setminus \{0\})$, of
compact support away from zero momentum. These $N$ photons are
prepared at time $t= -\tau$ in such a way that they hit the atom 
at rest at time $t=0$. (We thank S.~De Bi{\`e}vre, M.~Griesemer, H.~Spohn
and especially S.~Teufel for clarifying this point to us.)
{F}rom Eq.~(\ref{eq-II-6}) below we show that, in particular, 
$A(\infty, \uf) \gs := \lim_{\tau \to \infty} A(\tau, \uf) \gs$ exists,
thus representing an incoming photon scattering state.
In contrast, choosing $\Psi$ according to (\ref{eq-I-37}) 
and with $\tau=0$ amounts to adding at time $t=0$ the photon cloud
$A(\uf)$ to the (interacting) atom at rest. 

We henceforth often leave out trivial tensor factors in our notation
whenever it is clear from the context what is meant. For instance,
we write $A(\uf) = a^*(f_1) \, a^*(f_2) \, \cdots \, a^*(f_N)$,
$H_0 = \hel + \hf$, etc.

\subsubsection{The Photoelectric Effect}
\label{subsubsec-I-2-3}
Having defined the transported charges with momentum in
$\cT$, we can now formulate a quantitative assertion about it. 
To this end, we fix $p \in \RR^3$ and define two distributions 
$L_0^p, L_\infty^p \in \cD'(\RR^3 \setminus \{0\} )$ on smooth functions 
$f \in C_0^\infty( \RR^3 \setminus \{0\} )$ of compact support away 
from zero by 
\begin{eqnarray} \label{eq-I-41a} 
\la L_\infty^p \, , \, f \ra & := & 
2 \pi \int \delta\big( p^2 - E_0 - \om(k) \big) \, 
     \hrho(p, k) \, f(k) \, d^3k \comma
\\ \label{eq-I-41b}
\la L_0^p \, , \, f \ra & := &
\int \frac{ -i \, \hrho(p, k) \, f(k) \, d^3k }
          { p^2 - E_0 - \om(k) + i0 } 
\\ \label{eq-I-41c}
& = &
- \pi \int \delta\big( p^2 - E_0 - \om(k) \big) \, 
     \hrho(p, k) \, f(k) \, d^3k 
\\ \nonumber & & 
- \, i \, \PP \int \frac{ \hrho(p, k) \, f(k) \, d^3k }
                   { p^2 - E_0 - \om(k) }  \comma
\end{eqnarray}
$\PP \int $ denoting the Cauchy principal value and 
$\hrho(p,k)$ denoting the partial Fourier transform of 
$\rho(x,k)$ (as a square-integrable function) with respect
to the particle position variable $x$. 

Now we are in position to formulate our main result.
\begin{theorem} \label{thm-I-1}
Assume Hypotheses~\ref{H-1}--\ref{H-2}, fix $N,m \in \NN$ with
$m \leq N$, and let
$\vphi_1, \vphi_2, \ldots \vphi_m \in C_0^{\infty}(\RR^3 \setminus \{0\})$
be an orthonormal system, $\la \vphi_i | \vphi_j \ra = \delta_{i,j}$.
Let $A(\tau, \uvphi)$ be the corresponding photon cloud, i.e.,
\begin{equation} \label{eq-I-43} 
A(\tau, \uvphi) \ = \ 
e^{-i\tau H_g} \, e^{i\tau H_0} \: 
a^*(\vphi_1)^{n_1} \, a^*(\vphi_2)^{n_2} \; \cdots \: a^*(\vphi_m)^{n_m} 
\: e^{-i\tau H_0} \, e^{i\tau H_g}  \comma
\end{equation}
where $n_j \in \NN$ are such that $n_1 + n_2 + \ldots + n_m =N$,
and fix a measurable set $\cT \subseteq \RR^3$. 
\begin{itemize}
\item[(i)]
If $\tau > g^{-1/K}$ then the transported charges
with momentum in $\cT$ of the initial state $A(\tau, \uvphi) \gs$
satisfy
\begin{eqnarray} \label{eq-I-44a} 
\lefteqn{
\Qinf_\cT \big( A(\tau, \uvphi) \gs \big) \, + \, \cO\big( g^{2+\mu} \big) 
\ = \ 
\Qsup_\cT \big( A(\tau, \uvphi) \gs \big) \, + \, \cO\big( g^{2+\mu} \big) 
}
\\[1mm] \nonumber & & = \ \;
g^2 \, Q^{\tau}_\cT (\uvphi) \ \; := \ \;
g^2 \, \big( n_1! \, n_2! \cdots n_m! \big) \,
\sum_{j=1}^m n_j \, \int_\cT 
|\la L_\infty^p \, , \, \vphi_j \ra|^2 d^3p \period
\end{eqnarray}
\item[(ii)]
If $\tau < g^{\mu}$ then the transported charge
with momentum in $\cT$ of the initial state $A(\tau, \uvphi) \gs$ is given 
by
\begin{eqnarray} \label{eq-I-44b} 
\lefteqn{
\Qinf_\cT \big( A(\tau, \uvphi) \gs \big) \, + \, \cO\big( g^{2+\mu} \big) 
\ = \ 
\Qsup_\cT \big( A(\tau, \uvphi) \gs \big) \, + \, \cO\big( g^{2+\mu} \big) 
}
\\[1mm] \nonumber & & = \ \;
g^2 \, Q^{\tau}_\cT (\uvphi) \ \; := \ \;
g^2 \, \big( n_1! \, n_2! \cdots n_m! \big) \,
\sum_{j=1}^m n_j \, \int_\cT 
|\la L_0^p \, , \, \vphi_j \ra|^2 d^3p \period
\end{eqnarray}
\end{itemize}
Here, $L_\infty^p$ and $L_0^p$ are the distributions defined in
Eqs.~(\ref{eq-I-41a})--(\ref{eq-I-41c}), and  
$\mu \equiv \mu(K) := 1 - K^{-1} >0$, where $K$ is the degree
of differentiability in Hypothesis~\ref{H-1} and \ref{H-2}.  
\end{theorem}
 The asymptotic representation~\eqref{eq-I-44a}
(resp.~\eqref{eq-I-44b}) is similar to the Born expansion for the
$T$-matrix (see e.g.~\cite{MR80m:81085}) in Schr{\"o}dinger scattering
theory (resp. to the Born expansion for the wave operator). 

Note that the coefficient $Q^{\tau}_\cT (\uvphi)$ of the leading
order contribution to the photoelectric effect is \emph{additive}
(counting multiplicities) in the single photon states 
$\vphi_1, \ldots, \vphi_m$. This implies that the contribution of 
each $\vphi_j$ is independent of the other single photon states 
$\vphi_i$, $i \neq j$,
in the photon cloud, in contrast to the classical model for the
photoelectric effect which would have predicted a dependence of
$Q^{\tau}_\cT (\uvphi)$ only on the \emph{total energy} of the
photon cloud.

\subsubsection{The Limit of Monochromatic Light}
\label{subsubsec-I-2-4}
For large values of $\tau$, i.e., in Case~(i) in
Theorem~\ref{thm-I-1} above, the transported charge directly yields
Einstein's predictions (\ref{eq-I-1}) and (\ref{eq-I-2}). In Case~(ii),
however, there is an inconsistency between Eq.~(\ref{eq-I-44b})
and (\ref{eq-I-1}) and (\ref{eq-I-2}), since 
$\la L_0^p \, , \, \vphi_j \ra$ may be non-vanishing, even if 
$\vphi_j$ is supported in a region where $\om(k) < p^2 - E_0$.
This observation may reflect that Case~(ii) in Theorem~\ref{thm-I-1} 
is physically less relevant than Case~(i), because it can hardly 
be realized experimentally: the photons would need to be present
at the origin $x=0$ at time $t=0$, coming from nowhere.
On the other hand, Case~(ii) deals with a (perhaps naive, but reasonable) 
first proposal for a model of a photon cloud, and it is 
appropriate to discuss this case, as well.

It turns out, that Einstein's predictions (\ref{eq-I-1}) and
(\ref{eq-I-2}) can be recovered even in Case~(ii), provided the
incident light is sufficiently monochromatic, i.e., sharply localized
in momentum space. To define our notion of monochromatic light

in precise terms, we restrict
ourselves to considering a photon cloud consisting of single photon
with wave function $\vphi_\delta \in C_0^\infty(\RR^3\setminus\{0\})$,
only. It is a trivial matter to extend the consideration below to a
photon cloud of $N$ orthonormal photon wave functions.

For fixed $\om >0$, we construct a wave function $\vphi_\delta$
localized in energy about $\om$ by choosing a smooth,
$L^2(\RR)$-normalized function of one, compactly supported variable,
$\chi_1, \chi_2, \ldots \chi_m \in C_0^{\infty}[(-\frac{1}{2},\frac{1}{2})]$, 
and a smooth, normalized functions on the two-sphere, 
$\kappa \in C^{\infty}[S^2]$ and setting 
\begin{equation} \label{eq-I-46} 
\vphi_{\delta} (r,\Omega_k)
\ := \ \om^{-1} \, \delta^{-1/2} \,
\chi\Big( \frac{r - \omega_j}{\delta} \Big) \, \kappa(\Omega_k) \comma
\end{equation}
using spherical coordinates $(r,\Omega_k) = (|k|, k/|k|)$.
The limit of monochromatic light is then defined to be $\delta \to 0$.
Note that $\vphi_\delta$ is asymptotically normalized in this
limit, $\lim_{\delta \to 0} \|\vphi_\delta\| = 1$.

Given the photon wave function $\vphi_\delta$, a 
(more or less straightforward) computation in distribution theory
yields, for any measurable set $\cT \subseteq \R^3$, that
the first order term of the transported charge in Case~(ii)
in Theorem~\ref{thm-I-1} is given by
\begin{equation} \label{eq-I-51a}
\lim_{\delta \to 0} \int\limits_{\cT}  
|\la L_0^p \, , \, \varphi_{\delta} \ra |^2 \, d^3p 
\ = \ 
\left\{ \begin{array}{ccc}
                      0 & \mbox{if} & E_0 + \om < 0 \comma \\
\Theta_{0} \: I_{0} & \mbox{if} & E_0 + \om = 0 \comma \\ 
\Theta_{0} \: I_{+} & \mbox{if} & E_0 + \om > 0 \comma \\ 
\end{array} \right.
\end{equation}
where,
\begin{eqnarray} \label{eq-I-51b}
\Theta_{0} & := & \int_{\SS^2}
\one_{\cT} \big( \sqrt{E_0+\om} \sigma ) \;
\sqrt{E_0+\om \: } \; 
\big| \Theta(\sqrt{E_0+\om} \, \sigma \: , \: \om) \big|^2 \, 
d^2\sigma \comma \hspace{6mm}
\\ \label{eq-I-51bb}
\Theta(p,r) & := &  
-i \int_{\SS^2} \frac{r^2}{\om} \, \hrho(p, r \sigma) \, \kappa(\sigma)
\, d^2\sigma \comma \hspace{6mm}
\end{eqnarray}
with $\hrho(p, k)$ denoting the Fourier transform of $x \to \rho(x,k)$,
and
\begin{eqnarray} \label{eq-I-51c}
I_{0} & := & 
\int_0^\infty \bigg| \int_{-\infty}^\infty 
\frac{\chi(x) dx}{y-x+i0} \bigg|^2 \, dy \comma
\\ \label{eq-I-51d} 
I_{+} & := & 
\int_{-\infty}^\infty \bigg| \int_{-\infty}^\infty 
\frac{\chi(x) dx}{y-x+i0} \bigg|^2 \, dy \comma
\end{eqnarray}
We state Eq.~(\ref{eq-I-51a}) without proof but mention that its
derivation requires a slightly stronger assumption than what is
provided by Hypothesis~\ref{H-1}--\ref{H-2}.

Einstein's predictions (\ref{eq-I-1}) and (\ref{eq-I-2}) are now
manifest in the right hand side of (\ref{eq-I-51a}).  The
coefficient $\lim_{\delta \to 0} Q^{0}_\cT (A_\delta \gs)$ may still
vanish, as the factor $\Theta_{0}$ may turn out to be zero. 
This case is in accordance with (\ref{eq-I-1}) and
(\ref{eq-I-2}) and merely reflects the physical fact that some
transitions (here: from a bound into a scattering state) are forbidden
in lowest order perturbation theory.

\subsubsection{Comparison to other Results and Methods in Scattering Theory}
\label{subsubsec-I-2-5}

As pointed out in the introduction, the results of the present paper
may be considered scattering theoretic, as the transported charge
derives from the asymptotic limit of the unitary evolution
operator $e^{-itH_g}$, as $t \to \infty$. Most of the work in
mathematical scattering theory in the past two decades or so is
devoted to proving asymptotic completeness (AC), first for
nonrelativistic $N$-body systems and, more recently and inspired
by the former, for quantum field theoretic models like the one
defined by the Hamiltonian $H_g$, especially, in 
\cite{DerezinskiGerard1997,DerezinskiGerard1999,FroehlichGriesemerSchlein2001,FroehlichGriesemerSchlein2001a,Gerard1996}.
The mathematical methods developed and applied in these papers are, 
perhaps, more sophisticated than those used by us for the computation
of the transported charge. A central role in all the papers mentioned
above is played by \emph{asymptotic observables}, notably the
\emph{asymptotic velocity operator}. Loosely speaking, the positivity
of the asymptotic velocity guarantees that any initial state 
eventually breaks up into parts which move independently, i.e., whose
time evolution is free (noninteracting), and this is an important
ingredient to prove AC. It is therefore important to note that 
the transported charge measures \emph{any} contribution that escapes
to infinity (in particle configuration space), even those that
come from particles with zero asymptotic velocity. Another important
aspect to note is that we are interested in a \emph{quantitative}
estimate on the transported charge, asymptotically in $g$.
In this context, AC has to be seen as an existence result:
every vector can be approximated by a polynomial in 
asymptotic creation operators, acting on the ground states of the system.
It is not easy to turn the involved estimates leading to this 
statement into quantitative information. Also, it should be noted
that so far AC can be proved only for massive quantum fields --
an assumption that plays no role in our analysis, although we do
make a confinement assumption on the interaction couplings.

To be more specific about the implications of AC on the value of the 
transported charge, we consider the results in 
\cite{FroehlichGriesemerSchlein2001}, where AC is proved
for Rayleigh scattering, i.e., for the energy range below the 
ionization threshold. In \cite{FroehlichGriesemerSchlein2001} an
auxiliary Hilbert space $\widetilde{\cH}=\cH \otimes \cF$ is introduced,
and with the scattering identification operator 
$I:\widetilde{\cH} \longrightarrow \cH$
given by $I[\vphi \otimes a^*(f_1) \cdots a^*(f_n) \Omega] := 
a^*(f_1) \cdots a^*(f_n) \vphi$, the dynamics generated
by $\widetilde{H}:=H_g \otimes \one_{\cF} + \one_{\cH} \otimes \hf$ on
$\widetilde{\cH}$ can be compared to the dynamics on $\cH$ generated by
$H_g$. The main result in \cite{FroehlichGriesemerSchlein2001} is, that
\begin{equation} \label{eq-I-36e}
\Omega_+:=s-\lim_{\tau \to \infty} e^{i\tau H_g} I e^{-i\tau \widetilde{H}} 
(\one_{pp}(H_g) \otimes \one_{\cF})
\end{equation}
exists and $\Ran \Omega_+ \supseteq \Ran \one_{\RR_0^-}(H_g)$.
In particular, for each $\Psi \in \Ran \one_{\RR_0^-}(H_g)$, 
there exists $\vphi \in \widetilde{\cH}$, such that
\begin{eqnarray} \label{eq-I-36f} 
\lefteqn{
\limsup_{R \to \infty} \: \limsup_{t \to \infty} \: 
\| F_R \, e^{-itH_g} \, \Psi \|
\ = \
\limsup_{R \to \infty} \: \limsup_{t \to \infty} \: 
\| F_R \, e^{-itH_g} \, \Omega_+ \, \vphi \|
\ = \
} 
\nonumber \\
& = & 
\limsup_{R \to \infty} \:\limsup_{t \to \infty} \:
\|F_R \, I \, e^{-it\widetilde{H}} \, (\one_{pp}(H_g) \otimes \one_{\cF}) \, 
\vphi \| 
\ = \ 0 \comma \hspace{20mm}
\end{eqnarray}
and the last equation follows from the fact that 
$e^{-it\widetilde{H}}$ leaves $\one_{pp}(\cH)$ invariant, and the
application of $I$ does not affect the electronic component, thus 
(\ref{eq-B-1a}) implies (\ref{eq-I-36f}). Hence, the proof of AC
in \cite{FroehlichGriesemerSchlein2001} yields
$\Qinf = \Qsup = 0$ below the ionization threshold.
While this result does not cover the range of positive energies, 
which is where the photoelectric effect takes place,
the proof of AC for Compton scattering in the more recent paper
\cite{FroehlichGriesemerSchlein2001a} may show that 
$\Qinf = \Qsup$ extends to any bounded energy range.
It would then be interesting to see whether the methods used therein
also yield a better quantitative estimate on the transported charge
than the one derived in the present paper. 


\secct{Dyson Series and Photoelectric Effect} \label{sec-II}
%
In this section we study the asymptotics, as $t \to \infty$, for
$e^{-itH_g} A(\tau, \uf) \gs$, where $H_g = H_0 + gW$ is the Hamiltonian
generating the dynamics of the interacting system, which is applied
to an initial state $A(\tau, \uf) \gs$ for $0 \leq \tau \leq g^{\mu}$ and for
$\tau \geq g^{-\frac{1}{K}}$. The initial state is composed of
a photon cloud 
\begin{equation} \label{eq-II-1a} 
A  \equiv A(\uf) \: = \: a^*(f_1) a^*(f_2) \cdots  a^*(f_N)
\comma \hspace{2mm} \mbox{with} \hspace{2mm} 
f_1, f_2, \ldots, f_N \, \in \, C_0^\infty(B_\Lambda \setminus \{0\}) \comma 
\end{equation}
of $N$ photons and its time evolution 
\begin{equation} \label{eq-II-1b} 
A(\tau) \, \gs \ \equiv \ A(\tau,\uf) \gs \ := \ 
e^{-i\tau H_g} \, e^{i\tau H_0} \: A(\uf) \: 
e^{-i\tau H_0} \, e^{i\tau H_g} \, \gs
\end{equation}
applied to a ground state vector $\gs$,
\begin{equation} \label{eq-II-2}
H_g \, \gs \ = \ E_0 \, \gs \comma 
\end{equation}
representing the interacting atom at rest. Note that each photon
orbital $f_j$ is assumed to be smooth and compactly supported away
from zero momentum and in the ball of radius $\Lambda<\infty$ 
about the origin, 
\begin{equation} \label{eq-II-2.1}
\supp \, f_j \ \subseteq \ B_\Lambda \setminus \{0\} 
\ := \ \big\{ k \in \RR^3 \; \big| \; 0 < \om(k) < \Lambda \big\} \period
\end{equation}
We now give a list of assumptions used to derive Theorem~\ref{thm-II-1},
below. Given an operator $X$ on $\cH$ and 
$t \in \RR$ we denote its \emph{free} Heisenberg time evolution by
\begin{equation} \label{eq-II-3}
X_t \ := \ e^{-itH_0} \, X \, e^{itH_0} \period 
\end{equation}
Next, we recall from (\ref{eq-I-34a}) the estimate
\begin{equation} \label{eq-II-4} 
\big\| (\hfL+\one)^\alpha \, P_c \, \gs \big\| \: + \:
\big\| (\hfL+\one)^\alpha \, P_\Om^\perp \, \gs \big\| 
\ \leq \ \cO(g) \comma
\end{equation}
for any $\alpha \geq 1$ and any $\Lambda < \infty$, where
$\hfL$ is defined in (\ref{eq-I-34b}). Eq.~(\ref{eq-II-4}) is proved in
Lemma~\ref{lem-IV-1}, below. Furthermore, we recall from 
(\ref{eq-I-23.1}) that 
\begin{equation} \label{eq-II-5}
P_d \, W \, P_d 
\ \equiv \ 
(P_d \otimes \one) \; W \; (P_d \otimes \one) 
\ = \ 0 \period
\end{equation}
In Lemma~\ref{lem-IV-3}, we demonstrate that, for any $s \in \RR$,
\begin{equation} \label{eq-II-6}
\big\| \, [W \, , \, A_s] \, (\hfL + \one)^{-1 - (N/2)} \big\|
\ \leq \ \cO\big( (1 + |s|)^{-K} \big) \comma
\end{equation}
where $K >1$ is the degree of differentiability in Hypothesis~\ref{H-1}. 
Since $\gs \in \dom[(\hf+\one)^{1+(N/2)}]$, for any
$N \in \NN_0$, the estimate (\ref{eq-II-6}) and the fact that
$\hf$ and $P_d$ commute imply the norm-convergence of
\begin{equation} \label{eq-II-6.1}
\Phi_\infty \big( A(\uf), \tau \big) \, := \, 
\lim_{T \to \infty} \Phi_T \big( A(\uf), \tau \big) \comma
\end{equation}
where 
\begin{equation} \label{eq-IIa-6} 
\Phi_T(X, \tau) \, := \, \left \{ \begin{array}{lll} 
\displaystyle \int_0^T ds \, e^{is(H_0-E_0)} \, [W \, , \, X_s] 
\, P_d \, \gs  &
\mbox{if} & 0 \leq \tau \leq g^{\mu} \\[0.4cm] 
\displaystyle \int \limits_{-T}^T ds \, e^{is(H_0-E_0)} \, [W\, ,\, X_s]
\,  P_d \, \gs &
\mbox{if} & \tau \geq g^{-\frac{1}{K}} \end{array} \right.
\end{equation}
provided $K > 1$. Furthermore, we show in Lemma~\ref{lem-IV-4} that, 
for any $r \in \RR$ and $s \in \RR_0^+$,
\begin{equation} \label{eq-II-7}
\big\| \, W \, e^{-irH_0} \, P_c \, [W \, , \, A_s] 
       \, (\hf + \one)^{-2 - (N/2)} \big\|
\ \leq \ \cO\big( (1 + |r|)^{-3/2} \big) \period
\end{equation}
Under these assumptions, we prove the following Theorem.
\begin{theorem} \label{thm-II-1}
Assume Hypotheses~\ref{H-1}--\ref{H-2} and (\ref{eq-II-1a})--(\ref{eq-II-7}).
Then there exists a constant $C < \infty$ such that, 
for all $t \geq g^{-1/K}$,
\begin{equation} \label{eq-II-8}
\big\|
e^{-it(H_g-E_0)} \, A(\tau) \, \gs \; - \; A_t \, \gs \; + \;
      ig e^{-it(H_0-E_0)} \, \Phi_{\infty} (A, \tau) \big\| 
\ \leq \ C \, g^{1+\mu(K)} \comma
\end{equation}
provided that either $0 \leq \tau \leq g^{\mu}$ or 
$\tau \geq g^{-\frac{1}{K}}$, and
where $\mu(K) := 1 - K^{-1} > 0$ and $K>1$ 
is the degree of 
differentiability in Hypothesis~\ref{H-1}.
\end{theorem}
\proof\hfil
First we compute the Heisenberg time evolution of the photon cloud 
$A(\tau)$ on $\Ran$\newline\noindent $[(H_g-i)^{-1-N}]$,
\begin{eqnarray} \label{eq-II-10}
e^{-itH_g} \, A(\tau) \, e^{itH_g}
& = &
e^{-i(t+\tau)H_g} \, e^{i(t+\tau)H_0} \, A_t \, 
e^{-i(t+\tau)H_0} \, e^{i(t+\tau)H_g}
\\ \nonumber & = &
A_t \; + \; 
\Big[ e^{-isH_g} \, e^{isH_0} \, A_t 
\, e^{-isH_0} \, e^{isH_g} \Big]_{s=0}^{s=t+\tau}
\\ \nonumber & = &
A_t \; - \; i g \int_0^{t+\tau} ds \: 
e^{-isH_g} \, \big[ W \, , \, A_{t-s} \big] \, e^{isH_g} 
\\ \nonumber & = &
A_t \; - \; i g \int_{-\tau}^t ds \: 
e^{-i(t-s)H_g} \, \big[ W \, , \, A_s \big] \, e^{i(t-s)H_g} 
\comma
\end{eqnarray}
using $A = A(0) = A(0,\uf)$ and 
a change of integration variables, $s \mapsto t-s$.
Applying (\ref{eq-II-10}) to the ground state $\gs$, we obtain
\begin{eqnarray} \label{eq-II-13}
-ig \, \Psi(t,\tau) & := & 
e^{-it(H_g-E_0)} \, A(\tau) \, \gs \: - \: A_t \, \gs
\\ \nonumber & = & 
-ig \int_{-\tau}^t ds \, e^{-i(t-s)(H_g-E_0)} \, [W \, , \, A_s] \, \gs \comma
\end{eqnarray}
and our goal is to show that 
\begin{equation} \label{eq-II-14}
\big\| \Psi(t,\tau) \; - \; 
       e^{-it(H_0-E_0)} \, \Phi_\infty(A(\tau),\tau)) \big\| 
\ \leq \ C \, g^\mu \period
\end{equation}

To this end, we first prove (\ref{eq-II-14}) for the special case $\tau = 0$.
Given $T \geq 1$ and $K > 1$, and assuming that $t \geq T$, we observe that
\begin{eqnarray} \label{eq-II-15}
& \big\| \Phi_\infty (A,0) \; - \; \Phi_T(A,0) \big\|
   \ \leq \  \cO\big( T^{1-K} \big) \comma & 
\\[1mm] \label{eq-II-17}
& \bigg\| \Psi(t,0) \; - \;
\int_0^T ds \, e^{-i(t-s)(H_g-E_0)} \, [W \, , \, A_s] \, \gs \bigg\| 
\ \leq \ \cO\big( T^{1-K} \big)\period &
\end{eqnarray}
Next, we use that $H_0$ and $A$ commute with $P_d$ and
that $P_d \, W \, P_d =0$ which imply
\begin{equation} \label{eq-II-18}
[W \, , \, A_s] \, P_d \ = \ P_c \, [W \, , \, A_s] \, P_d \period
\end{equation}
Hence we have that
\begin{eqnarray} \label{eq-II-19}
\lefteqn{
\big\| e^{-itH_g} \, \Phi_T(A,0) \; - \; e^{-itH_0} \, \Phi_T(A,0) \big\|
}
\\ \nonumber & = &
g \bigg\| \int_0^t dr \, e^{-irH_g} \, W \, e^{-i(t-r)H_0} 
          \, \Phi_T(A,0) \bigg\| 
\\ \nonumber & = &
g \bigg\| \int_0^t dr \int_0^T ds \, e^{-isE_0} \, e^{-irH_g} \, W \, 
e^{-i(t-r-s)H_0} \, P_c [W \, , \, A_s] \, P_d \, \gs \bigg\|
\\ \nonumber & \leq &
g \big\| (\hf+\one)^{2+(N/2)} \gs \big\| 
\\ \nonumber & & \hspace{5mm}
\int_0^t dr \int_0^T ds \, 
\big\| W \, e^{-i(t-r-s)H_0} \, P_c [W \, , \, A_s] \, 
       (\hf+\one)^{-2-(N/2)} \big\| \comma
\end{eqnarray}
and (\ref{eq-II-7}) and a change of variable $r \mapsto t-r$ now imply that
\begin{eqnarray} \label{eq-II-20}
\big\| e^{-itH_g} \, \Phi_T(A,0) \; - \; e^{-itH_0} \, \Phi_T(A,0) \big\|
& \leq & 
C_1 \, g \, \int_0^T ds \int_0^t \frac{ dr }{ (1 + |r-s|)^{3/2} } 
\nonumber \\[1mm] & \leq & 
C_1 \, C_2 \, g \, T \comma
\end{eqnarray}
where $C_1 < \infty$ and 
$C_2 := \int_{-\infty}^\infty (1 + |r|)^{-3/2} dr < \infty$.
Applying the estimates (\ref{eq-II-15})--(\ref{eq-II-20}) to the
left side of (\ref{eq-II-14}), we hence obtain that
\begin{eqnarray} \label{eq-II-21}
\lefteqn{
\big\| \Psi(t,0) \; - \; e^{-it(H_0-E_0)} \, \Phi_\infty(A,0) \big\| 
}
\\ \nonumber & \leq & 
\Big\| \Psi(t,0) \, - \,  
\int_0^T ds \, e^{-i(t-s)(H_g-E_0)} \, [W \, , \, A_s] \, \gs \Big\|
\\ \nonumber & & 
\; + \;
\Big\| \int_0^T ds \, e^{-i(t-s)(H_g-E_0)} \, [W \, , \, A_s] \, \gs
\, - \, e^{-it(H_g-E_0)} \Phi_T(A,0) \big\| 
\\ \nonumber & & 
\: + \:
\big\| \big( e^{-it(H_g-E_0)} \, - \, e^{-it(H_0-E_0)} \big) 
                                                \Phi_T(A,0) \big\| 
\: + \: \big\|  \Phi_T(A,0) \, - \, \Phi_\infty(A,0) \big\| 
\\ \nonumber & \leq & 
\Big\| \int_0^T ds \, e^{is(H_g-E_0)} \, [W \, , \, A_s] \, \gs
\; - \; \Phi_T(A,0) \Big\| \; + \; 
\cO\big( T^{1-K} + g T \big) 
\\ \nonumber & \leq & 
\Big\| \int_0^T ds \, \big( e^{is(H_g-E_0)} - e^{is(H_0-E_0)} \big)
\, [W \, , \, A_s] \, P_d \, \gs \Big\| 
\\ \nonumber & & 
\; + \; \Big\| \int_0^T ds \, e^{is(H_g-E_0)} 
\, [W \, , \, A_s] \, P_c \, \gs \Big\|
\; + \; 
\cO\big( T^{1-K} + g T \big) \comma
\end{eqnarray}
using that $P_d + P_c = \one$. It remains to bound the two integrals
in the last line of (\ref{eq-II-21}). The second of these is bounded by
\begin{eqnarray} \label{eq-II-22}
\lefteqn{
\Big\| \int_0^T ds \, e^{is(H_g-E_0)} 
\, [W \, , \, A_s] \, P_c \, \gs \Big\|
}
\\ \nonumber & \leq & 
\big\| (\hfL+\one)^{1+(N/2)} \, P_c \, \gs \big\| \; 
\int_0^T ds \big\| \, [W \, , \, A_s] \, (\hfL + \one)^{-1 - (N/2)} \big\|
\\ \nonumber & \leq & 
\cO( g ) \ \leq \ \cO( g T ) \comma
\end{eqnarray}
due to (\ref{eq-II-4}) and (\ref{eq-II-6}), since $K > 1$ and $T \geq 1$.
To estimate the first integral, we use again the DuHamel formula
as in (\ref{eq-II-19}). This yields
\begin{eqnarray} \label{eq-II-23}
\lefteqn{
\Big\| \int_0^T ds \, \big( e^{is(H_g-E_0)} - e^{is(H_0-E_0)} \big)
\, [W \, , \, A_s] \, P_d \, \gs \Big\|
}
\\ \nonumber & \leq &
g \bigg\| \int_0^T ds \, e^{-isE_0} \int_0^s dr \, 
e^{irH_g} \, W \, e^{i(s-r)H_0} \, P_c [W \, , \, A_s] \, P_d \, \gs \bigg\|
\\ \nonumber &  \leq &
g \big\| (\hf+\one)^{2+(N/2)} \gs \big\| 
\\ \nonumber & & \hspace{5mm}
\int_0^T ds \int_0^s dr \, 
\big\|W e^{i(s-r)H_0} \, P_c [W \, , \, A_s] \, 
       (\hf+\one)^{-2-(N/2)} \big\| 
\\ \nonumber &  \leq &
C_1 \, g \int_0^T ds \int_0^s \frac{dr}{(1 + |r|)^{3/2}}
\ \leq \ \cO\big( g T \big) \comma
\end{eqnarray}
changing again the integration variable $r \mapsto s-r$.
Inserting (\ref{eq-II-22}) and (\ref{eq-II-23}) into  (\ref{eq-II-21}),
we arrive at 
\begin{equation} \label{eq-II-24}
\big\| \Psi(t,0) \; - \; e^{-it(H_0-E_0)} \, \Phi_\infty(A,0) \big\| 
\ \leq \ \cO\big( T^{1-K} + g T \big) \comma
\end{equation}
which yields (\ref{eq-II-14}) upon choosing $T := g^{\mu(K) -1}$ and setting 
$\mu(K): = 1 - K^{-1} >0$, for the special case $\tau = 0$.

Note that if more generally $0 \leq \tau < g^\mu$ then 
$\Phi_\infty(A,\tau) = \Phi_\infty(A,0)$ and hence
\begin{equation} \label{eq-IIa-1} 
\|A(\tau) \gs \, - \, A \gs\| \ = \ 
g \|\int_0^\tau ds \, e^{-is(H_g-E_0)} [W,A_{-s}]\gs\| \period
\end{equation}
Additionally using (\ref{eq-II-6}), we hence conclude
$\|A(\tau) \gs - A \gs \|\leq \cO(g^{1+\mu})$, 
proving the asymptotic expansion (\ref{eq-II-8}) in the case 
$\tau \in [0,g^{\mu}]$. 

In order to deal with the case $\tau \in [g^{-1/K}, \infty)$, we
observe that an application of
(\ref{eq-IIa-1}), for $t \geq g^{-\frac{1}{K}}$, in combination with 
the decay property (\ref{eq-II-6}) establishes
\begin{equation} \label{eq-IIa-2}
\| e^{-itH_g} A(t) \gs \, - \, e^{-itH_g} A(\tau) \gs\| 
\ \leq \ \cO(g^{1+\mu}) \comma
\end{equation}
and we may replace the time-evolved state $e^{itH_g}A(\tau) \gs$ 
by $e^{-itH_g}A(t) \gs$ in this case. Using $\tau = t$ in 
Eq.~(\ref{eq-II-10}), we furthermore obtain
\begin{equation} \label{eq-IIa-3}
e^{-it(H_g-E_0)} A(t) \gs \ = \ 
A_t \gs \; - \; ig \int_{-t}^t ds \, e^{-i(t-s)H_g} \, [W,A_s] \, 
e^{i(t-s)H_g} \gs \period
\end{equation}
Using
\begin{equation} \label{eq-IIa-4} 
\Psi(t,t) \ = \ \int_{-t}^t ds \, e^{-i(t-s)(H_g-E_0)} \, [W,A_s] \, \gs 
\comma 
\end{equation}
computations running along the lines of (\ref{eq-II-15})--(\ref{eq-II-23})
substituting 
$\Phi_{\infty}(A,\infty)$, $ \Phi_{T}(A,\infty)$, $ \Psi(t,\infty)$ for
$\Phi_{\infty}(A,0)$, $ \Phi_{T}(A,0)$, $ \Psi(t,0)$
and replacing integrals of the form 
$\int\limits_0^T ds$ by those of the form $\int\limits_{-T}^T ds$ proves
(\ref{eq-II-8}) for $\tau \geq g^{-\frac{1}{K}}$, as well. \qed

Theorem~\ref{thm-II-1} is the main tool for the analysis of the 
charges transported to infinity, $\Qinf_{\cT}(A(\tau)\gs)$ and 
$\Qsup_{\cT}(A(\tau)\gs)$,
because it allows us to take the limits $t \to \infty$ and $R \to \infty$,
as shown in the following lemma.
\begin{lemma} \label{lem-II-1}
Assume Hypotheses~\ref{H-1}--\ref{H-2}, and let
$A = a^*(f_1) \, a^*(f_2) \cdots$\newline\noindent $a^*(f_N)$ with  
$f_1,...,f_N \in C_0^{\infty}(\RR^3 \setminus \{0\})$. Then there exists
a constant $C < \infty$, such that the charges transported to infinity 
obey
\begin{equation} \label{eq-II-25}
\Qsup_{\cT}(A(\tau)\gs) \, - \, C g^{2+\mu} 
\ \leq \ g^2 \big\| T_{\cT} \, \Phi_{\infty}(A,\tau) \big\|^2
\ \leq \ 
\Qinf_{\cT}(A(\tau) \gs) \, + \, C g^{2+\mu} \comma
\end{equation}
for $0 \leq \tau \leq g^{\mu}$ and
$\tau \geq g^{-\frac{1}{K}}$, where $\mu(K) := 1 - K^{-1} > 0$ 
and $K>1$ is the degree of 
differentiability in Hypothesis~\ref{H-1}.
\end{lemma}
\Proof
First we observe that, since $F_R \equiv F_R \otimes \one_f$ commutes with 
$A_t \equiv \chf_{\el} \otimes A_t$ and
$H_f=\chf_{\el} \otimes H_f$, we have 
\begin{eqnarray} \label{eq-II-29}
\big\| T_{\cT} \, F_R \, A_t \, \gs \big\|
& \leq &
\big\| F_R \, A_t \, \gs \big\| 
\ \; = \ \; 
\big\| A_t \, F_R \, \gs \big\| 
\\ \nonumber  & \leq &
\big\| A_t \, (\hf+1)^{-N/2} \big\| \cdot 
\big\| F_R \, (\hf+1)^{N/2} \, \gs \big\| \period
\end{eqnarray}
Applying Lemma~\ref{lem-IV-2} $N$-times, we observe that
$\big\| A_t \, (\hf+1)^{-N/2} \big\|$ is bounded by
$C_2 < \infty$, uniformly in $t$,
\begin{equation} \label{eq-II-30}
\big\| A_t \, (\hf+1)^{-N/2} \big\| \ \leq \ 
\prod_{j=1}^N \Big\| (\hf+1)^{(j-1)/2} \, 
   a^*(e^{-it\om} f_j) \, (\hf+1)^{-j/2} \Big\| 
\ \leq \ C_2 \period
\end{equation}
Moreover, $(\hf+1)^{N/2} \gs \in \cH$ and 
$F_R \to 0$ strongly, as $R \to \infty$, hence
\begin{equation} \label{eq-II-31}
\lim_{R \to \infty} \lim_{t \to \infty}
\big\| T_{\cT} \, F_R \, A_t \, \gs \big\| \ = \ 0 \period
\end{equation}
According to Theorem~\ref{thm-II-1} we have
\begin{equation} \label{eq-II-26} 
e^{-it(H_g-E_0)} \, A(\tau) \, \gs 
\ = \ 
A_t \, \gs \: - \: ig \, e^{-it(H_0-E_0)} \, \Phi_{\infty}(A,\tau)
\: + \: \Rem(t) \comma
\end{equation}
where $\| \Rem(t) \| \leq C_1 g^{1+\mu}$, for some constant $C_1 < \infty$
independent of $t \in \RR^+$ and $\mu = 1 - 1/K$.
Inserting Eqs.~(\ref{eq-II-26}) and (\ref{eq-II-31}) into the definitions 
(\ref{eq-I-36a})--(\ref{eq-I-36b}) of the transported charges, we find
that
\begin{equation} \label{eq-II-27}
\Qsup_{\cT}(A(\tau)\gs) \, - \, C_1 g^{2+\mu} 
\ \leq \ g^2 \tQ_{\cT}(A(\tau)\gs)
\ \leq \ 
\Qinf_{\cT}(A(\tau) \gs) \, + \, C_1 g^{2+\mu} \comma
\end{equation}
where
\begin{equation} \label{eq-II-28}
\tQ_{\cT}(A(\tau)\gs) \ := \ 
\lim_{R \to \infty} \lim_{t \to \infty}
\big\| T_{\cT} F_R e^{-it(H_0-E_0)} \Phi_{\infty}(A,\tau) \big) \big\|^2 
\comma
\end{equation}
and estimate (\ref{eq-II-25}) would follow from 
$\tQ_{\cT}(A(\tau)\gs) = \big\| T_{\cT} \Phi_{\infty}(A,\tau) \big\|^2$.

Next, we turn to 
$T_{\cT} F_R e^{-it(H_0-E_0)} \Phi_{\infty}(A,\tau)$.
We note that $F_R = F_R \, P_c$ and hence
\begin{equation} \label{eq-II-32}
F_R \, e^{-it(H_0-E_0)} 
\ = \ 
e^{-it(H_f-E_0)} \, F_R \, P_c \, e^{-it\hel}
\ = \ 
e^{-it(H_f-E_0)} \, F_R \, e^{-it(-\Delta)} \, P_c \period
\end{equation}
Furthermore, the absolute continuity of the spectrum of $-\Delta$
on $L^2(\RR^3)$ implies that 
\begin{equation} \label{eq-II-33}
(1-F_R) e^{-it(-\Delta)} \ = \ \one_{\{|x| < R\}} e^{-it(-\Delta)} \to 0
\end{equation}
strongly, as $t \to \infty$, for any $R < \infty$. Using (\ref{eq-II-32}),
(\ref{eq-II-33}), and the fact that the operators $T_{\cT} = T_{\cT} P_c$, 
$e^{-it(-\Delta)}$, and $e^{-it(H_f-E_0)}$ commute, we thus obtain  
\begin{eqnarray}
\lefteqn{ \label{eq-II-34}
\lim_{R \to \infty} \lim_{t \to \infty}
\big\| T_{\cT} \, F_R \, e^{-it(H_0-E_0)} \, \Phi_{\infty}(A,\tau) \big\|=}\\
& = &
\lim_{R \to \infty} \lim_{t \to \infty}
\big\| T_{\cT} \, F_R \, 
e^{-it(-\Delta)} \, P_c \, \Phi_{\infty}(A,\tau) \big\|=
\big\| T_{\cT} \, \Phi_{\infty}(A,\tau) \big\|
\nonumber  \comma
\end{eqnarray}
which implies the claim.\QED

Next, we recall from Eq.~(\ref{eq-II-6.1}) the definition of
\begin{equation} \label{eq-II-35}
\Phi_\infty (A,\tau) \, = \,\left\{  \begin{array}{ll} 
\displaystyle \int_0^\infty ds \, e^{is(H_0-E_0)} \, [W \, , \, A_s] \, P_d \, \gs &
{\text{if }} 0 \leq \tau \leq g^{\mu} \\[0.4cm]
\displaystyle \int_{-\infty}^{\infty} ds \, e^{is(H_0-E_0)} \, [W \, , \, A_s] \, P_d \, 
\gs &
{\text{if }} \tau \geq g^{-\frac{1}{K}} \end{array} \right.
\end{equation}
which we complement by the definition of the vector
\begin{equation} \label{eq-II-36}
\Phi_\infty^\Om (A,\tau) \,: = \, 
\left\{  \begin{array}{ll} 
\displaystyle \int_0^\infty ds \, e^{is(H_0-E_0)} \, [W \, , \, A_s] \, P_d \,
(0 \oplus \Om) &
{\text{if }} 0 \leq \tau \leq g^{\mu} \\[0.4cm]
\displaystyle \int_{-\infty}^{\infty} ds \, e^{is(H_0-E_0)} \, [W \, , \, A_s] \, P_d \, (0 \oplus \Om) &
{\text{if }}\tau \geq g^{-\frac{1}{K}} \end{array} \right.
\end{equation}
whose existence is guaranteed thanks to Lemma~\ref{lem-IV-3}. 
Lemma~\ref{lem-II-2} below shows that these two vectors differ
by at most $\cO(g)$, i.e., a replacement of the ground state
$\gs$ of $H_g$ by the ground state $0 \oplus \Om$ of $H_0$
in the definition of $\Phi_\infty (A,\tau)$ introduces only small errors
which are negligible, as we shall see.
\begin{lemma} \label{lem-II-2}\hfil
Assume Hypotheses~\ref{H-1}--\ref{H-2}, and let
$A = a^*(f_1) \, a^*(f_2) \cdots a^*(f_N)$ with  
$f_1,...,$\newline\noindent$f_N \in C_0^{\infty}(\RR^3 \setminus \{0\})$. Then there exists
a constant $C < \infty$, such that 
\begin{equation} \label{eq-II-37}
\big\| \Phi_\infty (A,\tau) \, - \, \Phi_\infty^\Om (A,\tau) \big\|
\ \leq \ C \, g \period
\end{equation}
\end{lemma}
\Proof
We first note that due to Lemma~\ref{lem-IV-3}, we have the estimate
\begin{eqnarray} 
\lefteqn{ \label{eq-II-38}
\big\| \Phi_\infty (A,\tau) \, - \, \Phi_\infty^\Om (A,\tau) \big\|
\leq }\\ \nonumber
&\leq & \Big\| (\hfL + 1)^{1 + \frac{N}{2}} 
\big( P_d \gs \, - \, (0 \oplus \Om) \big) \Big\| 
\cdot \int_{\R} ds 
\big\| \, [W \, , \, A_s] \, (\hfL + 1)^{-1 - \frac{N}{2}} \big\|
\\ \nonumber & \leq & 
C_1 \, \Big\| (\hfL + 1)^{1 + \frac{N}{2}} 
\big( P_d \gs \, - \, (0 \oplus \Om) \big) \Big\| \comma
\end{eqnarray}
for some constant $C_1 < \infty$.
Additionally using Lemma~\ref{lem-IV-1} and the decomposition
$P_d \gs - (0 \oplus \Om) = \big[ P_d P_\Om \gs - (0 \oplus \Om) \big]
+ P_d P_\Om^\perp \gs$ , we observe that
\begin{equation} \label{eq-II-39}
\Big\| (\hfL + 1)^{1 + N/2} 
\big( P_d \gs \, - \, (0 \oplus \Om) \big) \Big\| 
\ \leq \ \cO(g) \period \hspace{40mm} \QED
\end{equation}

The final ingredient for the proof of Theorem~\ref{thm-I-1} is 
Lemma~\ref{lem-II-3}, below. Note that due to the definitions (\ref{eq-II-36})
we may restrict ourselves to the case $\tau =0$ and $\tau =\infty$.
\begin{lemma} \label{lem-II-3}
Assume Hypotheses~\ref{H-1}--\ref{H-2}, fix $N,m \in \NN$ with
$m \leq N$, and let
$\vphi_1, \vphi_2, \ldots \vphi_m \in C_0^{\infty}(\RR^3 \setminus \{0\})$
be an orthonormal system, $\la \vphi_i | \vphi_j \ra = \delta_{i,j}$.
\hfil Let $A = a^*(\vphi_1)^{n_1} a^*(\vphi_2)^{n_2} \cdots$\newline\noindent
$a^*(\vphi_m)^{n_m}$, 
where $n_j \in \NN$ are such that $n_1 + n_2 + \ldots + n_m =N$,
and fix a measurable set $\cT \subseteq \RR^3$. Then 
\begin{eqnarray} 
\label{eq-II-40} 
\big\| T_{\cT} \, \Phi_{\infty}^\Om (A,\tau) \big\|^2
& = & 
Q^{\tau}_\cT (\uvphi) = \\
& = & 
\big( n_1! \, n_2! \cdots n_m! \big) \,
\sum_{j=1}^m n_j \, \int_\cT 
|\la L_\tau^p \, , \, \vphi_j \ra|^2 d^3p \comma \nonumber
\end{eqnarray}
with $L_\tau^p = L_0^p$, for $\tau \in [0,g^\mu]$, and
$L_\tau^p = L_\infty^p$, for $\tau \in [g^{-1/K}, \infty)$,
where $L_0^p$ and $L_\infty^p$ are defined in 
Eqs.~(\ref{eq-I-41a})--(\ref{eq-I-41c}).
\end{lemma}
\Proof
We first write $W = a^*(G) + a(G)$ and use the canonical 
commutation relations to obtain that
\begin{eqnarray} \label{eq-II-41}
[ W , A_s ] 
& = & 
\big[ a^*(G) + a(G) \ , \ a^*(e^{-is\om} \vphi_1)^{n_1} 
      a^*(e^{-is\om} \vphi_2)^{n_2}
          \cdots  a^*(e^{-is\om} \vphi_m)^{n_m} \big]
\nonumber \\ & = &
\big[ a(G) \ , \ a^*(e^{-is\om} \vphi_1)^{n_1} 
      a^*(e^{-is\om} \vphi_2)^{n_2}
          \cdots  a^*(e^{-is\om} \vphi_m)^{n_m} \big]
\nonumber \\ & = &
\sum_{j=1}^m  n_j \: 
\bigg\{ \big\la G \big| e^{-is\om} \vphi_j \big\ra \otimes
\prod_{i=1}^m a^*(e^{-is\om} \vphi_i)^{n_i-\delta_{i,j}} \bigg\} \comma
\end{eqnarray}
on $\dom[\hf^{1+N/2}]$, where 
$\la G | e^{-is\om} \vphi_j \ra \in \cB(\cH_\el)$ is a bounded
operator acting on the electron variables given by
\begin{equation} \label{eq-II-42}
\big\la G \big| e^{-is\om} \vphi_j \big\ra
\ := \ 
\int d^3k \: e^{-is\om(k)} \: \vphi_j(k) \, G^*(k) \period 
\end{equation}
Moreover, for any $s \geq 0$,
\begin{eqnarray} \label{eq-II-43}
\lefteqn{
e^{is(H_0 - E_0)} \big\la G \big| e^{-is\om} \vphi_j \big\ra 
\otimes 
\prod_{i=1}^m a^*(e^{-is\om} \vphi_i)^{n_i-\delta_{i,j}} \, (0 \oplus \Om)
}
\nonumber \\ & = &
e^{is(H_\el - E_0)} \big\la G \big| e^{-is\om} \vphi_j \big\ra 
\otimes e^{is\hf}
\prod_{i=1}^m a^*(e^{-is\om} \vphi_i)^{n_i-\delta_{i,j}} \, (0 \oplus \Om) 
\nonumber \\ & = &
e^{is(-\Delta - E_0)} 
\big\la P_c \, G \, P_d \big| e^{-is\om} \vphi_j \big\ra  \otimes 
\prod_{i=1}^m a^*(\vphi_i)^{n_i-\delta_{i,j}} \, (0 \oplus \Om) \period
\nonumber \\ & = &
\big( \psi_{j,s} \oplus 0 \big) \otimes 
\prod_{i=1}^m a^*(\vphi_i)^{n_i-\delta_{i,j}} \Om \comma
\end{eqnarray}
where the Fourier transform of $\psi_j \in L^2(\RR^3)$ with
respect to the particle coordinate $x$ is given by
\begin{equation} \label{eq-II-44} 
\hpsi_{j,s}(p)
\ = \ 
\int d^3k \: e^{is(p^2 - E_0 - \om(k))} \: \hrho(p,k) \,
\vphi_j(k) \period 
\end{equation}
Here, we used the fact that $P_d \, G(k)^* P_d =0$
and hence $G(k)^* P_d = P_c \, G(k)^* P_d$, for all $k \in \RR^3$.
Moreover, $P_c \, G(k)^* P_d (\tpsi \oplus z) = z p_\downarrow(k)^* \oplus 0$,
and this yields (\ref{eq-II-44}).
Eqs.~(\ref{eq-II-41})--(\ref{eq-II-44}) imply that
\begin{equation} \label{eq-II-45} 
\Phi_{\infty}^\Om (A, \tau) 
\ = \ 
\sum_{j=1}^m n_j \, \bigg\{ \int_{-\tau}^\infty ds \, 
\big( \psi_{j,s} \oplus 0 \big) \bigg\}
\otimes
\prod_{i=1}^m a^*(\vphi_i)^{n_i-\delta_{i,j}} \Om \comma
\end{equation}
Passing to the momentum representation for the particle variable
and using that 
\begin{equation} \label{eq-II-46} 
\bigg\la \prod_{i=1}^m a^*(\vphi_i)^{n_i-\delta_{i,j}} \, \Om
\; \bigg| \;\; 
\prod_{i=1}^m a^*(\vphi_i)^{n_i-\delta_{i,\ell}} \, \Om \bigg\ra
\ = \ 
\delta_{j,\ell} \cdot \prod_{i=1}^m (n_i-\delta_{i,j})! \comma
\end{equation}
we hence obtain
\begin{eqnarray} \label{eq-II-47} 
\big\| T_{\cT} \, \Phi_{\infty}^\Om (A, \tau) \big\|^2
& = & 
\bigg\| \sum_{j=1}^m n_j \, \Big\{ \int_{-\tau}^\infty ds \, 
\big( T_{\cT} \, \psi_{j,s} \oplus 0 \big) \Big\} \otimes
\prod_{i=1}^m a^*(\vphi_i)^{n_i-\delta_{i,j}} \Om \bigg\|^2
\nonumber \\ & = &
n_1! n_2! \cdots n_m! \, \sum_{j=1}^m n_j
\int_\cT d^3p  \bigg| \int_{-\tau}^\infty ds \, \hpsi_{j,s}(p) \bigg|^2 \period
\end{eqnarray}
It remains to evaluate the integral $\int_{-\tau}^\infty ds \, \hpsi_{j,s}(p)$
in (\ref{eq-II-47}), which is an exercise in distribution theory.
To this end, we note that the $K>1$-fold partial differentiability of 
$\rho(x,k)$ with respect to $k$ in Hypothesis~\ref{H-1} implies that 
$\hrho(p, \, \cdot \, ) \vphi_j$ is two times partially differentiable 
with respect to $k$, and that its derivatives of order $\leq 2$
have bounded support away from $k=0$. Since the phase in $\hpsi_{j,s}(p)$
is non-stationary, away from $k=0$, two times integration by parts 
back and forth yields
\begin{eqnarray} \label{eq-II-48} 
\lefteqn{
\int_1^\infty ds \, \hpsi_{j,s}(p) 
\ \; = \ \; 
\int_1^\infty ds \, \int d^3k \: e^{is(p^2 - E_0 - \om(k))} \: \hrho(p,k) \,
\vphi_j(k) 
}
\nonumber \\ & = &
\int_1^\infty \frac{ds}{s^2} \, \int d^3k \: e^{is(p^2 - E_0 - \om(k))} \: 
\Big(i \nabla \cdot \frac{k}{|k|} \Big)^2 
\Big[ \hrho(p,k) \, \vphi_j(k) \Big] 
\nonumber \\ & = &
\lim_{\eps \to 0} \int_1^\infty \frac{ds}{s^2} \, \int d^3k \: 
e^{is(p^2 - E_0 - \om(k) + i\eps)} \: 
\Big(i \nabla \cdot \frac{k}{|k|} \Big)^2 
\Big[ \hrho(p,k) \, \vphi_j(k) \Big] 
\nonumber \\ & = &
\lim_{\eps \to 0} \int d^3k \, \int_1^\infty ds \,  
e^{is(p^2 - E_0 - \om(k) + i\eps)} \: \hrho(p,k) \, \vphi_j(k) 
\nonumber \\ & = &
\lim_{\eps \to 0} \int d^3k \, 
\frac{-i \, e^{i(p^2 - E_0 - \om(k))} \, \hrho(p,k)}
     {p^2 - E_0 - \om(k) + i\eps} \, \vphi_j(k) \period
\end{eqnarray}
Thus in the case $\tau =0$:
\begin{eqnarray} \label{eq-II-49} 
\int_0^\infty ds \, \hpsi_{j,s}(p) 
& = & 
\int_0^1 ds \, \hpsi_{j,s}(p) \; + \;
\int_1^\infty ds \, \hpsi_{j,s}(p) 
\\ \nonumber & = &
\lim_{\eps \to 0} \int d^3k \, \frac{-i \, \hrho(p,k)}
{p^2 - E_0 - \om(k) + i\eps} \: \vphi_j(k) =
\big\la L_0^p \, , \, \vphi_j \big\ra 
\period 
\end{eqnarray}
For the case $\tau= \infty$, we get
\begin{eqnarray} \label{eq-IIa-8}
\lim_{t \to \infty} \int_{-t}^t ds \, \hpsi_{j,s}(p) 
& = & \lim_{ t \to \infty}
\int_{-t}^t ds \, \int d^3k \: e^{is(p^2 - E_0 - \om(k))} \: \hrho(p,k) \,
\vphi_j(k) =\\
& = & \lim_{ t \to \infty}
\int d^3k \: \frac{e^{it(p^2 - E_0 - \om(k))}-e^{it(p^2 - E_0 - \om(k))}}
{i(p^2-E_0-\omega(k))}  \: \hrho(p,k) \, \vphi_j(k) \nonumber \\
& = & \lim_{ t \to \infty} 2
\int d^3k \: \frac{\sin(t(p^2 - E_0 - \om(k)))}
{p^2-E_0-\omega(k)}  \: \hrho(p,k) \, \vphi_j(k) =\nonumber \\
&=& 2\pi \int_{\SS^2} d^2 \Omega (p^2-E_0)^2 \hat{\rho}(p;p^2-E_0,\Omega)
\varphi_j(p^2-E_0,\Omega) \nonumber .  \QED 
\end{eqnarray}

\noindent \textbf{Proof of Theorem~\ref{thm-I-1}:}
The proof of Theorem~\ref{thm-I-1} consists of the following chain of
estimates,
\begin{eqnarray} 
\Qinf_{\cT}(A(\tau)\gs) \, + \, \cO( g^{2+\mu} ) 
& = & 
\Qsup_{\cT}(A(\tau)\gs) \, + \, \cO( g^{2+\mu} ) 
\nonumber 
\\ \label{eq-II-50} & = & 
g^2 \big\| T_{\cT} \, \Phi_{\infty}(A,\tau) \big\|^2
\\ \label{eq-II-51} & = & 
g^2 \big\| T_{\cT} \, \Phi_{\infty}^\Om(A,\tau) \big\|^2 \, + \, 
\cO( g^{2+\mu} ) 
\\ \label{eq-II-52} & = & 
Q^{\tau}_\cT (\uvphi) \, + \, \cO( g^{2+\mu} ) \comma
\end{eqnarray}
where Lemma~\ref{lem-II-1}, \ref{lem-II-2}, and \ref{lem-II-3} 
justify Eq.~(\ref{eq-II-50}), (\ref{eq-II-51}), and (\ref{eq-II-52}), 
re\-spec\-ti\-ve\-ly. \QED


\secct{Technical Estimates} \label{sec-IV}
%
In this section we derive the estimates (\ref{eq-II-4}),
(\ref{eq-II-6}), and (\ref{eq-II-7}), which is the basic input
for the asymptotics of the time evolution asserted in 
Theorem~\ref{thm-II-1}. 

Before we turn to these bounds, we derive a preparatory lemma.
\begin{lemma} \label{lem-IV-2}
Suppose that $\alpha \geq 1$, $\Lambda <\infty$, and let
$G \in L^2[ \RR^3 ; \cB(\cH)]$ with 
$C := \int d^3k \big( 1 + \om(k)^{-1} \big) \, \|G(k)\|^2 \, < \infty$.
Denote $a^*(G) := \int d^3k \, G(k) \otimes a^*(k)$ and
$a(G) := \int d^3k \, G(k)^* \otimes a(k)$. Then 
\begin{eqnarray} 
\label{eq-IV-5} 
\big\| (\hfL+1)^{\alpha} \, a(G) \, (\hf+1)^{-\alpha - 1/2} \big\| 
& \leq & \sqrt{C} \comma 
\\ \label{eq-IV-6} 
\big\| (\hfL+1)^{\alpha} \, a^*(G) \, (\hf+1)^{-\alpha - 1/2} \big\| 
& \leq & \sqrt{C} \, (2+4\Lambda)^\alpha \comma 
\\ \label{eq-IV-7} 
\big\| (\hfL+1)^{\alpha} \, a(G_\Lambda) \, 
(\hfL+1)^{-\alpha - 1/2} \big\| 
& \leq & \sqrt{C} \comma 
\\ \label{eq-IV-8} 
\big\| (\hfL+1)^{\alpha} \, a^*(G_\Lambda) \, 
(\hfL+1)^{-\alpha - 1/2} \big\| 
& \leq & \sqrt{C} \, (2+4\Lambda)^\alpha \comma 
\end{eqnarray}
where $G_\Lambda(k) := \chf_{ \{\om(k) < \Lambda\} } \, G(k)$, 
and $\chf_{ \{\om(k) < \Lambda\} }$ denotes the characteristic function
of $\{ k \in \RR^3 \; | \; \om(k) < \Lambda \}$.
\end{lemma}
\proof
First, we introduce a more compact notation,
\begin{equation} \label{eq-IV-9} 
\begin{array}{lclclclclcl}
G_1 & := & G , & \ \ & H_1 & := & \hf, & \ \ & \om_1 & := & \om, \\
G_2 & := & G_\Lambda, & \ \ &  H_2 & := & \hfL, 
                      & \ \ & \om_2 & := & \om_\Lambda, \\
\end{array}
\end{equation}
and observe that Eqs.~(\ref{eq-IV-5})--(\ref{eq-IV-8}) are equivalent to
\begin{eqnarray} \label{eq-IV-10} 
\big\| (\hfL+1)^{\alpha} \, a(G_j) \, (H_j+1)^{-\alpha - 1/2} \big\| 
& \leq & \sqrt{C} \comma 
\\ \label{eq-IV-11} 
\big\| (\hfL+1)^{\alpha} \, a^*(G_j) \, (H_j+1)^{-\alpha - 1/2} \big\| 
& \leq & \sqrt{C} \, (2+4\Lambda)^\alpha \comma 
\end{eqnarray}
with $j=1,2$. We apply the operator on the left side of
(\ref{eq-IV-10}) to a normalized vector $\psi \in \cH$ and obtain the
desired estimate by means of the pull-through formula, the
Cauchy-Schwarz inequality, and $\hfL+1 \leq H_j+1+\om_j$,

\begin{eqnarray} \label{eq-IV-14} 
\lefteqn{
\big\| (\hfL+1)^{\alpha} \, a(G_j) \, (H_j+1)^{-\alpha -1/2} \, \psi \big\| 
}
\\ \nonumber & \leq &
\int d^3k \; \| G_j(k) \| \,
\bigg\| \frac{\hfL+1}{H_j+1+\om_j(k)} \bigg\|^\alpha \,
\Big\| a(k) \, ( H_j+1 )^{-1/2} \, \psi \Big\|
\\ \nonumber & \leq &
\bigg( \int \, \frac{ \| G_j(k) \|^2 \, d^3k}{ \om_j(k) } \bigg)^{1/2} \cdot
\bigg\| \frac{ H_j }{ H_j+1 } \bigg\|^{1/2}
\ \leq \ 
\bigg( \int \, \frac{ \| G_j(k) \|^2 \, d^3k}{ \om_j(k) } \bigg)^{1/2} \period
\end{eqnarray}
for any $\psi \in \cH$, $\|\psi\| =1$.
To derive (\ref{eq-IV-11}), we use the canonical commutation relations,
\begin{eqnarray} \label{eq-IV-15} 
\lefteqn{
\big\| (\hfL+1)^{\alpha} \, a^*(G_j) \, 
                (H_j+1)^{-\alpha - 1/2} \, \psi \big\|^2
}
\\ \nonumber & = &
\int d^3k \, d^3\tk \, 
\Big\la \psi \, \Big| \;
G_j(k)^* G_j(\tk) \otimes 
(H_j+1)^{-\alpha - 1/2} \, a(k) \, \big( \hfL + 1 \big)^{2\alpha}
\: a^*(\tk) \, 
\\ \nonumber & & \hspace{35mm}
(H_j+1)^{-\alpha - 1/2} \, \psi \Big\ra  
\\ \nonumber & = &
\int d^3k \, d^3\tk \,
\bigg\la a(\tk) \, (H_j+1)^{-1/2} \psi \bigg| \; G_j(k)^* G_j(\tk) 
\\ \nonumber & & 
\hspace{10mm} \otimes 
\bigg[\frac{ \big( \hfL + 1 + \om_\Lambda(k) + \om_\Lambda(\tk) \big)^2 }
     { \big( H_j + 1 + \om_j(k) \big) \, \big( H_j + 1 + \om_j(\tk) \big) } 
\bigg]^\alpha \: a(k) \, (H_j+1)^{-1/2} \, \psi \bigg\ra  
\\ \nonumber &  & + \;
\int d^3k \, 
\Big\la \psi \Big| \;
G_j(k)^* G_j(k) \otimes (H_j+1)^{-2\alpha -1} \, 
\big(\hfL + 1 + \om_\Lambda(k) \big)^{2\alpha} \, \psi \Big\ra  
\\ \nonumber & \leq &
\bigg\| \frac{ \hfL+1+2\Lambda }{ H_j+1 } \bigg\|^{2\alpha} \,
\bigg( \int d^3k \; \| G_j(k) \| \,
\Big\| a(k) \, ( H_j+1 )^{-1/2} \, \psi \Big\| \bigg)^2
\\ \nonumber &  & + \;
\Big\| (H_j+1)^{-2\alpha -1} \, 
\big(\hfL + 1 + \om_\Lambda(k) \big)^{2\alpha} \Big\| \:
\bigg( \int d^3k \, \| G_j(k) \|^2  \bigg)
\\ \nonumber & \leq &
C \, (1+2\Lambda)^{2\alpha} \, 
\bigg(1\: + \: \int d^3k \, \om_j(k) \,
\Big\| a(k) \, ( H_j+1 )^{-1/2} \, \psi \Big\|^2 \bigg)
\\ \nonumber & \leq &
C \, (1+2\Lambda)^{2\alpha} \,
\bigg(1\: + \: \Big\| \frac{H_j}{H_j+1} \Big\| \bigg) 
\ \leq \ C \, (2+4\Lambda)^{2\alpha} \period
\end{eqnarray}
Note that in the last step we used 
$H_j = \int d^3k \, \om_j(k) \, a^*(k) a(k)$.\QED

Now we are in position to establish Eq.~(\ref{eq-II-4})
which shows that the interacting atom-photon ground state $\gs$
is well-localized in energy, even for the noninteracting Hamiltonian.
\begin{lemma} \label{lem-IV-1}
For any $\alpha \geq 1$, $\Lambda <\infty$, and sufficiently small
$g>0$, we have
\begin{equation} \label{eq-IV-1} 
\big\| (\hfL+\one)^\alpha \, P_c \, \gs \big\| \: + \:
\big\| (\hfL+\one)^\alpha \, P_\Om^\perp \, \gs \big\| 
\ \leq \ \cO(g) \period
\end{equation}
\end{lemma}
\proof
The proof is similar to the one for \cite[Thm.~X]{BachFroehlichSigal1998a}.
We first note that the asserted bound (\ref{eq-IV-1}) is implied
by (\ref{eq-I-34}) and the following estimate,
\begin{equation} \label{eq-IV-1.1} 
\big\| (\hfL+\one)^\alpha \, \chf_{ \{ H_0 \geq E_0/3 \} } \, \gs \big\| 
\ \leq \ \cO(g) \comma
\end{equation}
because on $\Ran \, \chf_{ \{ H_0 < E_0/3 \} }$, we have $P_c =0$
and $\hfL < |e_0| - |E_0|/3$.

Let $\chi \in C_0^\infty( I ; [0,1] )$ be a real-valued, smooth 
function, compactly supported in 
$I := (\frac{4}{3} E_0 \, , \, \frac{2}{3} E_0 )$ and such that
$\chi(E_0) = 1$. Clearly, $\chf_{ \{ H_0 \geq E_0/3 \} } \, \chi(H_0) =0$,
and $\chi(H_g) \, \gs = \gs$. Hence we observe that
\begin{eqnarray} \label{eq-IV-2} 
\lefteqn{
(\hfL+1)^\alpha \, \chf_{ \{ H_0 \geq E_0/3 \} } \, \gs \ = \
}
\\ \nonumber & & 
\chf_{ \{ H_0 \geq E_0/3 \} } \, (\hfL+1)^\alpha \, 
\big[ \chi(H_g) - \chi(H_0) \big] \, 
(H_g - E_0 + 1)^{-\alpha -1} \gs \period
\end{eqnarray}
Next, we represent $\chi(H_g)$ and $\chi(H_0)$ by the functional
calculus based on almost analytic extensions of smooth functions
of compact support \cite{MelinSjoestrand1975}.
Let $\tchi \in C_0^\infty( \Delta; \CC)$ be an almost analytic extension
of $\chi$, supported in a small
complex neighborhood $\Delta \subseteq \CC$ of $I$ and such that
$\bar{\partial} \tchi(z) = \cO\big( (\rIm \, z)^2 \big)$.
By means of the measure 
$d\mu(z) := (2\pi i)^{-1} \bar{\partial} \tchi(z) \, dz \wedge d\bar{z}$ 
and the second resolvent equation, we have the following
identity
\begin{eqnarray} \label{eq-IV-3} 
\chi(H_g) - \chi(H_0) 
& = & 
\int \frac{d\mu(z)}{H_g - z} \: - \:  \frac{d\mu(z)}{H_0 - z}
\\ \nonumber & = &
- g \int d\mu(z) \big\{ (H_0 - z)^{-1} \, W \, (H_g - z)^{-1} \big\} \period
\end{eqnarray}
Inserting this identity into (\ref{eq-IV-2}) and applying
Lemma~\ref{lem-IV-2} to $W = a^*(G) + a(G)$, we obtain the norm
estimate
\begin{eqnarray} \label{eq-IV-4} 
\lefteqn{
\big\| (\hfL+1)^\alpha \, \chf_{ \{ H_0 \geq E_0/3 \} } \, \gs \big\|
}
\nonumber \\ & \leq &
g \, \bigg\| \int d\mu(z) \big\{ (H_0 - z)^{-1} \,
(\hfL+1)^{\alpha} \, W \,
\nonumber \\ & & \hspace{50mm} 
\cdot (H_g - E_0 + 1)^{-\alpha -1} \, (H_g - z)^{-1} \big\} \gs \bigg\|
\nonumber \\ & \leq &
g \, \int |d\mu(z)| \Big\{ \big\| (H_0 - z)^{-1} \big\| \,
\big\| (\hfL+1)^{\alpha} \, W \, (\hf+1)^{-\alpha - 1} \big\| \, 
\nonumber \\ & & \hspace{25mm} 
\cdot \big\| (\hf+1)^{\alpha + 1} \, 
(H_g - E_0 + 1)^{-\alpha -1} \big\| \, 
\big\| (H_g - z)^{-1} \big\| \Big\}
\nonumber \\ & \leq &
\cO(g) \comma
\end{eqnarray}
which finishes the proof.\QED

Our next goal is the derivation of (\ref{eq-II-6}). 
Actually, we prove a somewhat stronger estimate which is
also used in the proof of Lemma~\ref{lem-IV-4}, below.
Recall that
$A \equiv A(\uf) = a^*(f_1) a^*(f_2)$ $\cdots  a^*(f_N)$ is a photon
cloud of $N$ photons 
$f_1, f_2, \ldots, f_N \, \in \, C_0^\infty(B_\Lambda \setminus \{0\})$,
that $X_t :=  e^{-itH_0}  X  e^{itH_0}$ denotes the free
time evolution of an observable $X$, and that 
$K >1$ is the degree of differentiability in Hypothesis~\ref{H-1}. 
Denoting
\begin{equation} \label{eq-IV-15.1}
(1+|x|)^{3/2} \ := \ 
\left( \begin{array}{cc} 
(1+|x|)^{3/2} & 0 \\ 
      0       & 1 \\ 
\end{array} \right) \otimes \one \comma
\end{equation}
we prove the following lemma.
\begin{lemma} \label{lem-IV-3}
For any $N\in \NN$ and $K>1$, there exists a constant $C < \infty$
such that
\begin{equation} \label{eq-IV-16}
\big\| (1+|x|)^{3/2} \, [W \, , \, A_s] \, (\hfL + \one)^{-1 - (N/2)} \big\|
\ \leq \ C \, (1 + |s|)^{-K}  \comma
\end{equation}
for all $s \in \RR$.
\end{lemma}
\proof
First observe that on $\dom(\hf^{N/2})$, we have
$e^{-isH_0}  a^*(f)  e^{isH_0} = a^*(e^{-is\om} f)$ and hence
\begin{equation} \label{eq-IV-17}
A_s \ = \ a^*(e^{-is\om} f_1) \, a^*(e^{-is\om} f_2)
          \cdots  a^*(e^{-is\om} f_N) \period
\end{equation}
Writing $W = a^*(G) + a(G)$ and using the canonical commutation relations, 
we thus have
\begin{eqnarray} \label{eq-IV-18}
[ W \, , \, A_s ] 
& = & 
\big[ a^*(G) + a(G) \ , \ a^*(e^{-is\om} f_1) \, a^*(e^{-is\om} f_2)
          \cdots  a^*(e^{-is\om} f_N) \, \big]
\nonumber \\ & = &
\big[ a(G) \ , \ a^*(e^{-is\om} f_1) \, a^*(e^{-is\om} f_2)
          \cdots  a^*(e^{-is\om} f_N) \, \big]
\\ \nonumber & = &
\sum_{j=1}^N  \bigg\{ \big\la G \big| e^{-is\om} f_j \big\ra \otimes
\prod_{i=1, \atop i \neq j}^N a^*(e^{-is\om} f_i) \bigg\} \comma
\end{eqnarray}
where $\la G | e^{-is\om} f_j \ra \in \cB(\cH_\el)$ is a bounded
operator acting on the electron variables given by
\begin{equation} \label{eq-IV-19}
\big\la G \big| e^{-is\om} f_j \big\ra
\ := \ 
\int d^3k \: e^{-is\om(k)} \: f_j(k) \, G^*(k) \period 
\end{equation}
Note that $k \mapsto f_j(k) G^*(k)$ is $K >1$ times 
continuously differentiable, and, thanks to the support properties
of $f_j$, there exists $r_j >0$ such that
$f_j G^* \in C^K[ B_\Lambda \setminus B_{r_j} ; \cB(\cH_\el) ]$
has compact support away from zero. Thus, thanks to
Hypothesis~\ref{H-1}, $K$ times ($\cB(\cH_\el)$-valued)
integration by parts yields the standard estimate for oscillatory
integrals,
\begin{eqnarray} \label{eq-IV-20}
\lefteqn{
\Big\| (1+|x|)^{3/2} \, \big\la G \big| e^{-is\om} f_j \big\ra \, \Big\|
}
\\ \nonumber & \leq & 
|s|^{-K} \, \int d^3k \, 
\Big\| (1+|x|)^{3/2} \,
       \big( \nabla_k \cdot (k/|k|) \big)^K \: f_j(k) \, G^*(k) \Big\|
\nonumber \\ & \leq & 
\cO\big( (1+|s|)^{-K} \big) \comma
\end{eqnarray}
provided $|s| \geq 1$. For $|s| \leq 1$, Estimate~(\ref{eq-IV-20}) is 
trivial and hence holds for all $s \in \RR$.
Inserting (\ref{eq-IV-20}) into (\ref{eq-IV-18}) and undoing the
free time evolution (which is possible because $H_0$ commutes
with $\hfL$), we obtain
\begin{eqnarray} \label{eq-IV-21}
\lefteqn{
\big\| \, (1+|x|)^{3/2} \, [W \, , \, A_s] \, (\hfL + \one)^{-1 - (N/2)} \big\|
}
\\ \nonumber & \leq &
\cO\big( (1+|s|)^{-K} \big) \, 
\sum_{j=1}^N  \bigg\| \bigg( \prod_{i=1, \atop i \neq j}^N 
a^*(e^{-is\om} f_i) \bigg) \, (\hfL + \one)^{-1 - (N/2)} \bigg\| 
\\ \nonumber & = &
\cO\big( (1+|s|)^{-K} \big) \, 
\sum_{j=1}^N  \bigg\| \bigg( \prod_{i=1, \atop i \neq j}^N 
a^*(f_i) \bigg) \, (\hfL + \one)^{-1 - (N/2)} \bigg\| \period
\end{eqnarray}
To estimate a product of $N-1$ creation operators\hfil  
$a^*(\tf_1) a^*(\tf_2) \cdots  a^*(\tf_{N-1})$ with
$\tf_1, \tf_2, \ldots, $\newline\noindent$\tf_N \, \in \, 
  C_0^\infty(B_\Lambda \setminus \{0\})$,
we use Estimate~(\ref{eq-IV-8}) $N-1$ times and derive
\begin{eqnarray} \label{eq-IV-22}
\lefteqn{
\bigg\| \bigg( \prod_{n=1}^{N-1} 
a^*(\tf_n) \bigg) \, (\hfL + \one)^{-(N-1)/2)} \bigg\|
}
\\ \nonumber & = &
\bigg\| \prod_{n=1}^{N-1} \bigg( 
 (\hfL + \one)^{(n-1)/2} a^*(\tf_n) \, (\hfL + \one)^{-n/2} \bigg) \bigg\|
\\ \nonumber & \leq &
\prod_{n=1}^{N-1} \Big\| 
 (\hfL + \one)^{(n-1)/2} a^*(\tf_n) \, (\hfL + \one)^{-n/2} \Big\|
\ \leq \ \cO\big( (1+\Lambda)^{(N^2)} \big) \period
\end{eqnarray}
Inserting this bound into (\ref{eq-IV-21}) yields 
(\ref{eq-IV-16}), since $\Lambda < \infty$ is bounded.\QED

We finally turn to proving (\ref{eq-II-7}). Note that the following
Lemma uses $\hf$ rather than $\hfL$.
\begin{lemma} \label{lem-IV-4}
There exists a constant $C < \infty$ such that, 
for any $r \in \RR$ and $s \in \RR_0^+$,
\begin{equation} \label{eq-IV-23}
\big\| \, W \, e^{-irH_0} \, P_c \, [W \, , \, A_s] 
       \, (\hf + \one)^{-2 - (N/2)} \big\|
\ \leq \ C \, \big( (1 + |r|)^{-3/2} \big) \period
\end{equation}
\end{lemma}
\proof  \hfil
The proof is based on the decay of the Schwartz kernel 
$e^{-ir(-\Delta)}(x,y) = (4\pi i r)^{-3/2} \,$ \newline\noindent$ 
\exp[ i (x-y)^2 (4r)^{-1}]$
of the propagator $e^{-ir(-\Delta)}$ of the free particle.
More precisely, defining 
$C_1 := (4\pi)^{-1} \int (1+|x|)^{2\gamma} d^3x  < \infty$, 
we observe that
\begin{eqnarray} \label{eq-IV-24}
\lefteqn{
\big| \big\la (1+|x|)^{- \gamma} \vphi \; \big| \ 
      e^{-ir(-\Delta)} \, (1+|x|)^{-\gamma} \psi \big\ra \big|
}
\\ \nonumber & = &
\frac{1}{(4\pi \, r)^{3/2}}
\bigg| \int d^3x \, d^3y \, 
\frac{ \overline{\vphi(x)} \: e^{i(x-y)^2 / (4r)} \: \psi(y) }
     { (1+|x|)^{\gamma} \; (1+|y|)^{\gamma} } \bigg|
\\ \nonumber & \leq &
\frac{\|\vphi\| \; \|\psi\|}{(4\pi \, r)^{3/2}} \:
\bigg( \int \frac{d^3x}{ (1+|x|)^{2\gamma} } \bigg) 
\ \; \leq \ \; C_1 \, \|\vphi\| \, \|\psi\| \, r^{-3/2} \comma
\end{eqnarray}
for any $\vphi, \psi \in L^1 \cap L^2(\RR^3)$, since $\gamma > 3/2$.
This estimate yields
\begin{eqnarray} \label{eq-IV-25}
\lefteqn{
\big\| (1+|x|)^{-\gamma} \, e^{-irH_0} \, P_c \, (1+|x|)^{-\gamma} \big\|
}
\\ \nonumber & = & 
\left\| \left( \begin{array}{cc} 
(1+|x|)^{- \gamma} \, e^{-ir(-\Delta)} \, (1+|x|)^{-\gamma} & 0 \\ 
      0                                               & 0 \\ 
\end{array} \right) \otimes e^{-ir\hf} \right\|
\ \leq \ 
C_1 \, r^{-3/2} \period
\end{eqnarray}
Inserting (\ref{eq-IV-25}) into (\ref{eq-IV-23}) and using
that $\hf$ and $H_0$ commute, we obtain
\begin{eqnarray} \label{eq-IV-26}
\lefteqn{
\big\| \, W \, e^{-irH_0} \, P_c \, [W \, , \, A_s] 
       \, (\hf + \one)^{-2 - (N/2)} \big\|
}
\\ \nonumber & \leq &
\big\| W \, (1+|x|)^\gamma \otimes (\hf +1)^{-1/2} \big\| \;
\big\| (1+|x|)^{-\gamma} \, e^{-irH_0} \, P_c \, (1+|x|)^{-\gamma} \big\| \;
\\ \nonumber & & \hspace{20mm}
\big\| (1+|x|)^\gamma \otimes (\hf +1)^{1/2} \, 
       [W \, , \, A_s] \, (\hf + 1)^{-2 - (N/2)} \big\|
\\ \nonumber & \leq &
 C_1 \, r^{-3/2} \; 
\big\| W \, (1+|x|)^\gamma \otimes (\hf +1)^{-1/2} \big\| \;
\\ \nonumber & & \hspace{20mm}
\big\| (1+|x|)^\gamma \otimes (\hf +1)^{1/2} \, 
       [W \, , \, A_s] \, (\hf + 1)^{-2 - (N/2)} \big\| \period
\end{eqnarray}
Next, we use (\ref{eq-IV-18}) and the pull-through formula to commute
$(\hf +1)^{1/2}$ through $[W \, , \, A_s]$, observing that 
$\supp f_j \subseteq B_\Lambda(0)$. This yields
\begin{eqnarray} \label{eq-IV-27}
\lefteqn{
\Big\| (1+|x|)^\gamma \otimes (\hf +1)^{1/2} \, 
       [W \, , \, A_s] \, (\hf + 1)^{-2 - (N/2)} \Big\| \period
}
\\ \nonumber & \leq &
\Big\| (1+|x|)^\gamma \, [W \, , \, A_s] \, (\hf + 1)^{-2 - (N/2)} 
     \big(\hf +1 +(N-1)\Lambda \big)^{1/2} \Big\|
\\ \nonumber & \leq &
(1 + N\Lambda)^{1/2} \: \Big\| (1+|x|)^\gamma \, 
[W \, , \, A_s] \, (\hf + 1)^{-3/2 - (N/2)} \, \Big\|
\ \leq \ C_2 \comma
\end{eqnarray}
for some constant $C_2 < \infty$, uniformly in $s \in \RR$.
Here we additionally inserted (\ref{eq-IV-16}) to derive the last 
inequality. Finally, there exists a constant $C_3 < \infty$, such that
\begin{equation} \label{eq-IV-28}
\Big\| W \, (1+|x|)^\gamma \otimes (\hf +1)^{-1/2} \Big\|
\ \leq \ C_3 \comma
\end{equation}
as follows from writing $W \, (1+|x|)^\gamma =
a^*[ G \, (1+|x|)^\gamma] + a[ G \, (1+|x|)^\gamma]$
and applying Lemma~\ref{lem-IV-2}.\QED  

\newpage

\appendix

\secct{Elimination of Matrix Elements by a Bogoliubov Transformation}
\label{sec-A}
Consider $\hel = -\Delta - V(x)$, a Schr{\"o}dinger operator on $\RR^3$
having a single, simple negative eigenvalue, $e_0 <0$, associated to a
normalized vector $\vphi_0$. This Hamiltonian represents an atom
which we couple to the photonic field described in 
Subsect.~\ref{subsubsec-I-1-2} with an
interaction described in Subsect.~\ref{subsubsec-I-1-3}. 
More precisely, we assume the interaction $W$ to be of the
form (\ref{eq-I-13}), with $G \in L^2[ \RR^3; \cB(\cH_\el)]$ being a
square-integrable function with values in the bounded operators on
$\cH_\el$, satisfying 
\begin{equation} \label{eq-A-0-1}
\int d^3k \, \Big\{ 1 + \om(k)^{-2} \Big\} \, \big\| G(k) \big\|^2
\ \leq \ 1 \period
\end{equation}
Note that this assumption is stronger than (\ref{eq-I-14.3}).
We now show
\begin{theorem} \label{thm-A-1}
Assume (\ref{eq-A-0-1}).
For $g > 0$ small enough, there exists a unitary transform 
$U_g \in \cB(\cH_\el \otimes \cF)$ real analytic in $g$ such that
\begin{equation} \label{eq-A-1}
\tH_g  \ := \ U_g \, H_g \, U_g^*
\ = \thel \otimes \one_f \, + \, \one_\el  \otimes \hf  \, + \, g \tW
\end{equation}
where
\begin{itemize}
\item 
$\thel = \hel + g^2 \DV$, and $\DV$ is a bounded self-adjoint operator;
\item 
if $W$ acts as a multiplication operator in the electron variable,
$\DV$ is a potential, i.e., a multiplication operator;
\item 
$\thel$ has a single, simple negative eigenvalue, $e_g < 0$,
associated to the one-dimensional eigenprojector $\Pi_g$;
\item 
we have
\begin{equation} \label{eq-A-2}
(\Pi_g \otimes \one_f )\; \tW \; (\Pi_g \otimes \one_f) \ = \ 0 \period   
\end{equation}
\end{itemize}
All the quantities introduced above are real-analytic in $g$.
\end{theorem}
Theorem~\ref{thm-A-1} proves that, starting from a Schr{\"o}dinger
operator on $\RR^3$ with a single, simple negative bound state, using a
unitary transform, one can always pass to an interaction fulfilling
Assumption~(\ref{eq-I-14}).

\proof
Pick $h\in L^2(\R^3)$, and consider the Bogoliubov transform
\begin{equation} \label{eq-A-3}
U(h) \ = \ \one \otimes e^{ig[a^*(h) + a(h)]} \period
\end{equation}
A standard computation gives
\begin{equation} \label{eq-A-4}
U(h) \ H_g \ U(h)^* \ = \ 
\thel \otimes \one_f \; + \; \one_\el \otimes \hf \; + \; g \tW \comma
\end{equation}
where
\begin{eqnarray} \label{eq-A-5}
\thel & = &  \hel \; + \; g^2 \, \DV(h) \comma
\\ \label{eq-A-5-1}
\DV(h) & := & 
\int d^3k \Big\{ h(k)G(k) + \overline{h(k)}G^*(k) \Big\} \comma
\\ \label{eq-A-5-2}
\tW & = & W \; + \;  
\one \otimes \big\{ a^*(\om h) + a(\om h) \big\} \period
\end{eqnarray}
We remark that $\DV: L^2(\R^3) \to \cB(\cH_\el)$ is a bounded linear
operator with norm bounded by one, thanks to (\ref{eq-A-0-1}). Hence, 
$\thel$ has a single, simple eigenvalue $\te(h) = e_0 + \cO(g^2)$
in a vicinity of $e_0$, and we denote
the corresponding normalized eigenvector by $\vphi_g(h)$. 

To prove Theorem~\ref{thm-A-1}, we only need to show that, for $g>0$
sufficiently small, we can construct $h_g \in L^2(\R^3)$ such that
\begin{equation} \label{eq-A-6} 
- \la \vphi_g(h_g) , \hG(k) \vphi_g(h_g) \ra \ = \ h_g(k) \comma
\hspace{5mm} \mbox{ where } \hspace{5mm} 
\hG(k) \ := \ \om(k)^{-1} \, G(k) \period
\end{equation}
By Assumption~(\ref{eq-A-0-1}), $\hG$ is a
square-integrable function with values in the bounded operators on
$\cH_\el$. 
Using standard perturbation theory, we construct $\vphi_g(h)$ simply by
normalizing the vector 
\begin{equation} \label{eq-A-8}
\hvphi_g(h) 
\ := \ \frac{1}{2i\pi} \int_{\gamma_0}
\big(z - \hel - g^2 \DV(h) \big)^{-1} \, \vphi_0 \, dz \comma
\end{equation}
where $\gamma_0$ is a small circle of center $e_0$ and radius $|e_0/2|$.
The function $\vphi_g(h)$ is real analytic in $g$ and continuous in $h$, for
$h$ in the unit ball in $L^2(\RR^3)$; real analytic perturbation theory and
the bound $\| \DV(h) \| \leq \|h\|$ immediately give
\begin{equation} \label{eq-A-9}
\big\| \vphi_g(h) \, - \, \vphi_g(h') \big\| 
\ \leq \ C \, g^2 \, \| h-h' \| \comma
\end{equation}
for some constant $C<\infty$ and all $h, h' \in L^2(\RR^3)$
with $\|h\|,\|h'\|\leq 1$. Set
\begin{equation} \label{eq-A-10}
h_0 \ = \ - \la \vphi_0 | \, \hG(k) \, \vphi_0\ra \period
\end{equation}
For $g$ fixed and small and $h\in L^2(\R^3)$, with
$\| h \| \leq 1$, we define
\begin{equation} \label{eq-A-11}
T_g(h) \ = \ 
-\la \vphi_g( h_0 + h) \, | \, \hG(\cdot) \, \vphi_g(h_0+h) \ra 
\; - \; h_0 \period
\end{equation}
Note that $T_0$ has a trivial fixed point, $h=0$.
For $g>0$ sufficiently small, 
$T_g$ maps the unit ball of $L^2(\R^3)$ into itself. Moreover, one computes
\begin{eqnarray} \label{eq-A-12}
\big\| T_g(h)-T_g(h') \big\|
& \leq & 
4 \| \hG(k) \| \, \|\vphi_g(h_0+h)-\vphi_g(h_0+h') \|
\nonumber \\
& \leq & 
C \, g^2 \, \|h-h'\| \comma
\end{eqnarray}
for some $C<\infty$.
Hence, $T_g$ is contracting for $g$ small enough. 
Therefore, the fixed point equation
$h=T_g(h)$ has a unique solution in the unit ball of $L^2(\R^3)$.
This fixed point is the desired function solving (\ref{eq-A-6}).
We may construct this solution as the norm
limit of the sequence $T_g^n(0)$. Each of these terms being real
analytic in $g$ and the convergence being uniform in $g$ sufficiently
small (the rate of convergence is given by $g^2$), 
the limit is real analytic in $g$. \QED

\secct{Transported Charge for Bound and Negative Energy States}
\label{sec-B}
%

\begin{theorem} \label{thm-B-1}
Assume Hypothesis~\ref{H-1}. Then
\begin{eqnarray} \label{eq-B-1a} 
\lim_{R \to \infty} \: \sup_{t \geq 0} \,
\big\| F_R \: e^{-it H_g} \: 
   \: \one_{pp}(H_g) \: \Psi \, \big\| &=& 0
\\ \label{eq-B-1b} 
\lim_{R \to \infty} \: \sup_{t \geq 0} \, 
\big\| F_R \: e^{-it H_g} \:
\: \one_{]-\infty,0]}(H_g) \: \Psi \big\| \,&=&0
\end{eqnarray}
for any $\Psi \in \cH$, 
where $\one_{pp}(H_g)$ and $\one_{]-\infty,0]}(H_g)$ are the spectral
projections of $H_g$ onto its point spectrum and onto $]-\infty,0]$,
respectively.
\end{theorem}
\proof
To prove  equality  (\ref{eq-B-1a}), 
we first assume that $\Psi$ is an eigenvector of $H_g$
with corresponding eigenvalue $E \in \RR$. Since $F_R \to 0$ strongly,
as $R \to \infty$, we then have
\begin{equation} \label{eq-B-2} 
\lim_{R \to \infty} \, \sup_{t \geq 0}
\big\| F_R \:e^{-itH_g}\, \Psi \, \big\| =
\lim_{R \to \infty} \|F_R \Psi\|
\ = \ 0 \period
\end{equation}
This, of course, generalizes to any finite linear combination of
eigenvectors of $H_g$. Since the norm closure of this set of vectors
is the pure point subspace of $H_g$, Eq.~(\ref{eq-B-2}) also
generalizes to vectors in $\Ran \, \one_{pp}(H_g)$, which yields 
(\ref{eq-B-1a}). As a consequence of (\ref{eq-B-1a})
we may assume $\Psi = \one_c(H_g) \Psi$ in the proof of (\ref{eq-B-1b}). 
These vectors out of the continuous subspace have the property that 
\begin{equation} \label{eq-B-4} 
\lim_{\delta \searrow 0} \, \one_{(E-\delta, E+\delta)}(H_g) \Psi 
\ = \ 0 \comma
\end{equation}
for any $E \in \RR$. As $H_g$ is bounded from below and in view of 
(\ref{eq-B-4}), it hence (choosing $E:=0$) suffices to prove that
\begin{equation} \label{eq-B-5} 
\lim_{R \to \infty} \, 
\big\| F_R \: \chi[H_g] \, \big\| 
\ = \ 0 \comma
\end{equation}
for any smooth function $\chi \in C_0^\infty[ (-\infty, 0) ]$, 
compactly supported on the negative half-axis and away from zero.
The basic idea of our proof of (\ref{eq-B-5}) is essentially the same as
\cite[Thm.~II.1]{BachFroehlichSigal1998a} or Lemma~\ref{lem-IV-1}
and uses a Combes-Thomas or Agmon Estimate.
For $H = H_g$ or $H = H_0$, we use a representation 
\begin{equation} \label{eq-B-6} 
\chi[H] \ = \ 
\int_\Delta \frac{d\mu(z)}{H-z} \comma
\end{equation}
based on an almost analytic extension $\tchi \in C_0^\infty( \cM; \CC)$
of $\chi$, whose compact support $\cM \subseteq \CC$ 
can be chosen to include only $z \in \CC$ with $\rRe z \leq - \delta$,
for some $\delta >0$. Moreover, we can choose $\tchi$ as to obey
$\bar{\partial} \tchi(z) = \cO\big( (\rIm \, z)^2 \big)$.
The measure in (\ref{eq-B-6}) is then defined as
$d\mu(z) := (2\pi i)^{-1} \bar{\partial} \tchi(z) \, dz \wedge d\bar{z}$. 
By means of (\ref{eq-B-6}), the second resolvent equation, and the fact that
$F_R \, \chi[H_0] = F_R \, P_c \, \chi[H_0] =0$,
since $H_0 \geq 0$ on $\Ran \, P_c$, we thus have
\begin{eqnarray} \label{eq-B-7} 
F_R \, \chi[H_g] 
& = & 
F_R \, \Big( \chi[H_g] \, - \, \chi[H_0] \Big)
\nonumber \\ & = &
- g \int_\cM d\mu(z) F_R \, 
\frac{P_c}{-\Delta + H_f - z} \, W \, \frac{1}{H_g - z} \period
\end{eqnarray}
Next, we pick $1 < \lambda < \infty$ and introduce 
$f_\lambda :\RR^3 \to \RR^+$ by 
$f_\lambda(x) := \sqrt{ 1 + (x/\lambda)^2 \, }$,
and we denote $f_\lambda(x)^{-1} := 1/f_\lambda(x)$. 
We note the following properties of $f_\lambda$,
\begin{equation} \label{eq-B-8} 
\frac{|x|}{\lambda} \; \leq \; f_\lambda(x) \; \leq \; 1 + |x|^{3/2} 
\hspace{4mm} \mbox{and} \hspace{4mm}
f_\lambda(x)^{-1} \, |\nabla f_\lambda (x)|
\; \leq \; \lambda^{-1} \comma
\end{equation}
which imply 
\begin{eqnarray} 
\label{eq-B-9} 
\big\| F_R \, f_\lambda(x)^{-1} \big\| 
& \leq & C \, \lambda \, R^{-1} \comma 
\\ \label{eq-B-10}
\rRe\big\{ f_\lambda(x) \, (-\Delta) \, f_\lambda(x)^{-1} \big\}
& = & 
\\ \nonumber
-\Delta + 2 \rRe\big\{ 
f_\lambda(x)^{-1} \, \nabla f_\lambda (x) \cdot i\nabla \big\}
& \geq & - \lambda^{-2} \comma
\\ \label{eq-B-11}
\big\| f_\lambda(x) \, W \, (H_g + i)^{-1} \big\|
& \leq & C \comma
\end{eqnarray}
for some constant $C < \infty$. Here, (\ref{eq-B-9}) is trivial,
(\ref{eq-B-10}) is meant in the quadratic form sense on smooth
functions of compact support, and (\ref{eq-B-11}) uses
Hypothesis~\ref{H-1} and 
Lemma~\ref{lem-IV-2}. Choosing $\lambda \geq 2/\sqrt{\delta}$ and
inserting the bounds (\ref{eq-B-9})--(\ref{eq-B-11}) into
(\ref{eq-B-7}), we arrive at (\ref{eq-B-5}).\QED


\end{document}